\documentclass[10pt]{article}
\usepackage{amsmath,bbold,graphicx,array,slashed,amssymb,multirow,cite,eufrak}
\usepackage{pstricks}
\usepackage{color}
\newtheorem{theorem}{Theorem}

\begin{document}

\begin{titlepage}
\begin{center}
\vspace{1in}
\large{\bf Holographic Corrections to Meson Scattering Amplitudes}\\
\vspace{0.4in}
\large{Adi Armoni, Edwin Ireson}\\
\vspace{0.2in}
\emph{Department of Physics, Swansea University}\\ 
\emph{Singleton Park, Swansea, SA2 8PP, UK}\\
\vspace{0.3in}
\end{center}

\begin{abstract}
We compute meson scattering amplitudes using the holographic duality between confining gauge theories and string theory, in order to consider holographic corrections to the Veneziano amplitude and associated higher-point functions. The generic nature of such computations is explained, thanks to the well-understood nature of confining string backgrounds, and two different examples of the calculation in given backgrounds are used to illustrate the details. The effect we discover, whilst only qualitative, is re-obtainable in many such examples, in four-point but also higher point amplitudes.
\end{abstract}

\end{titlepage}

\tableofcontents

\newpage

\section{Introduction}

Calculating meson scattering amplitudes is a very hard problem, since it involves the strong coupling regime of QCD. Flat space string theory can be used to address this problem, but with partial success. Viewing the mesons as open strings in flat space leads to the Veneziano amplitude \cite{Veneziano:1968yb}, which admits certain appealing properties, mainly in the so-called ``Regge regime''. It is less successful in other kinematical regimes, and besides, flat space string theory is unrealistic- much work has been done since then, with curved string backgrounds and the holographic duality \cite{Maldacena:1997re}, to find more realistic physical phenomena within string theory.  Subsequently, the main purpose of this paper is to use modern string theory and the holographic duality to improve the results of the Veneziano amplitude by adding corrections due to holographic coordinates.

A few years ago Makeenko and Olesen \cite{Makeenko:2009rf} presented a derivation
 of the Veneziano amplitude from large-$N$ QCD, by using an (unrealistic) assumption that \textit{all} Wilson loops (large and small) admit an area law. Their derivation relies on the worldline formalism, where it can be shown that in the 't Hooft limit the meson scattering amplitude is given by a sum over all sizes and shapes of Wilson loops. A similar string theoretical derivation was presented recently in \cite{Armoni:2015nja}, where the Wilson loops are expressed as minimal surfaces in a holographic dual. The Veneziano amplitude is obtained by assuming that the string worlsheets accumulate on the end of space or horizon (the IR cutoff) and do not fluctuate, producing the same uniformly area-law behaved set of Wilson loops that Makeenko and Olesen formulated, thus naturally leading to the same result. But, from this starting point, it was also proposed that corrections to the Veneziano amplitude can be obtained by considering fluctuations in the vicinity of the IR cut-off of the dual holographic background.

The first step towards this goal was made recently in \cite{Armoni:2016nzm}. We considered meson scattering in Witten's background and studied the correction to the Veneziano amplitude due to the interaction of the 4d coordinates $X^\mu$ with the holographic coordinates. We found a deviation from the linear Regge trajectory near the origin. Such a deviation is expected from QCD, since a perfectly linear trajectory means a perfectly linear interquark potential (or an area law even for small Wilson loops). The perturbative regime of QCD tells us that the interquark potential admits a Coloumb-like behaviour and that small Wilson loops do not admit an area law. 

In this paper we elaborate and extend our previous results. Starting from large-$N$ QCD and using the worldline formalism, we derive an expression for the meson scattering amplitude in terms of a sum over string worldsheets that pass via the positions of the mesons. The worldsheets admit the topology of a disk. The boundary of the disk rests on the boundary of the AdS space and extends to the bulk. A typical contribution is depicted in fig.\ref{disk} below.
\begin{figure}[h]
\centering
\includegraphics[width=0.6\textwidth]{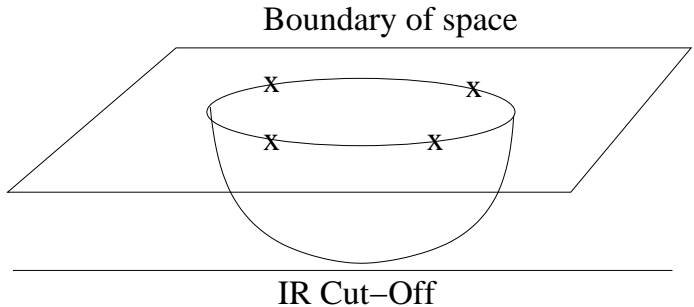}
\caption{A typical contribution to the meson scattering amplitude. A string worldsheet with a disk topology. The boundary of the disk sits on the boundary of the AdS space. The disk extends to the bulk and terminates on the IR cut-off. The locations of the mesons are denoted by x.}
\label{disk}
\end{figure}

The expression produces the Veneziano amplitude when the space is flat. In the general case, namely when the space is not flat, our expression should be viewed as the holographic prescription for calculating meson scattering.

We will argue that for a generic confining background the interaction of the 4d flat space with the extra coordinates is universal. More precisely, we argue that the worldsheet Lagrangian takes the following schematic form
\begin{equation}
 {\cal L}= \partial X^\mu \partial X_\mu +  \partial U \partial U
 + U^2\partial X^\mu \partial X_\mu \, ,
\end{equation}
namely it contains an interaction of the 4d coordinates with the holographic coordinates via a $U^2\partial X^\mu \partial X_\mu$ term. The same conclusion had already been made in \cite{Bigazzi:2004ze}. A perturbative correction to the $X^\mu$ propagator yields the correction to the scattering amplitude.

Apart from discussing the general case, we demonstrate our idea using two backgrounds: Witten's background of compactified D4 branes and the Klebanov-Strassler background. We also discuss the holographic correction to the multi-meson amplitude. 

There were previous studies of holographic corrections to hadron scattering \cite{Andreev:2004sy,Polchinski:2001tt}. The novelty of our approach is the use of $\alpha '$ corrections due to the interaction of 4d space with holographic coordinates.

The paper is organised as follows: in section 2 we derive our prescription for calculating meson scattering amplitudes. In section 3 we discuss the relation between the holographic criteria for confinement and the spaces that we will use. We then derive the generic form of the sigma models that we will use. In section 4 we focus on Witten's model and derive in great detail the correction to the Veneziano amplitude.  In section 5 we repeat the discussion for the Klebanov-Strassler case. In section 6 we discuss higher point functions. Section 7 is devoted to conclusions.

\section{Worldine formalism, Wilson loops and holography}

\subsection{The worldline approach to QCD}

To set up a map between our problem of computing meson operator expectation values and a more string-oriented calculation, we will employ the Worldline formalism. This procedure is essentially a rewriting of the QCD action as a sum of the expectation values of Wilson loops of all shapes and sizes. It is useful to us as these Wilson loop expectation values are objects whose holographic duals are well-understood: they are related to the action of a classical string hanging from the loop traced on the boundary of the holographic dual space, probing the bulk of the manifold. In particular, this worldline approach does away with the fermion determinant in the path integration, always a problematic term.

The approach is the following: the fermion determinant inside the path integral gets rewritten

\begin{align}
\mathcal{Z}=\int DA \exp(-S_{\rm YM}) \left( {\det} \left( i \slashed{D} \right) \right) ^{N_f}, \,\,\,\,\left( {\det} \left(i \slashed{D} \right)\right)  ^{N_f} &= \exp\left(-\frac{N_f}{2}\text{Tr}\int_0^\infty \frac{dT}{T} \mathcal{W}_T[A]\right)\text{ , }
\\ 
\mathcal{W}_T[A]=\int DxD\psi \exp\left( {-\frac{1}{2}\int_0^T d\tau\left(  \dot{x}^\mu \dot{x}_\mu + \psi^\mu  \dot{\psi}_\mu\right)}\right)  &\exp\left( i {\int_0^T d\tau\left( \dot{x}^\mu A_\mu - \frac{1}{2}\psi^\mu F_{\mu\nu} \psi^\mu\right)} \right) .\nonumber
\end{align}

One recognises the last expression as the expectation of a pure Yang-Mills Wilson loop of length $T$ and of arbitrary shape. We will take a large $N_c$ limit of the set-up, such that $N_f$ is kept constant, given that in general a product of $k$ connected Wilson loops scales as $1/N_c^{k-2}$, this makes the exponent small, and we will keep the linear term only.

We can use this prescription in any operator expectation value, in particular for the meson 4-point function $\left<\prod\limits_{i=1}^4 q\bar{q}(x_i)\right>$. For this purpose, we note that by turning on a current term in the action of the form $\int J \bar{q} q$, the partition function becomes
\begin{equation}
\mathcal{Z}[J]=\int DA \exp(-S_{\rm YM}) \left( {\det} \left( i \slashed{D} + J \right) \right) ^{N_f},
\end{equation}
from which the required expectation value becomes (again to linear order in our large $N_c$ expansion)
\begin{equation}
\left<\prod\limits_{i=1}^4 q\bar{q}(x_i)\right>=\frac{1}{\mathcal{Z}}\prod\limits_{i=1}^{4}\dfrac{\delta}{\delta J(x_i)}\mathcal{Z}[J]\\
=\int DA\,\,e^{-S_{\rm YM}} \left(-\frac{N_f}{2}\text{Tr}\int_0^\infty \frac{dT}{T}  \mathcal{W}_T[A] {\big|}_{x_1,x_2,x_3,x_4}\right),
\end{equation}
where the notation implies that these Wilson loops now always pass through the 4 relevant points in question thanks to the previous manipulation.

The gauge/gravity duality implies that, for the field theory on the boundary, a generic Wilson loop's expectation value is related to the expectation value of a string worldsheet hanging from the loop on the boundary down into the bulk. Thus we are able to write the following expression for the amplitude, suitably transformed out of position space into momentum space:

\begin{align}
\mathcal{A}\left(k_{i=1\dots4} \right) =\oint \prod\limits_{i=1}^4d\sigma_i \int [DX] W(\sigma_i)\exp\left( {ik_i^\mu X_\mu(\sigma_i)}\right)  \exp\left( -{1\over 2\pi \alpha '}{\int d^2\sigma\, G_{MN} \partial_\alpha X^M \partial ^\alpha X^N}\right) 
,\label{stringamp}
\end{align}
Where $W\exp\left( {ik_i^\mu X_\mu(\sigma_i)}\right)$ is the required meson operator in the spectrum of the (\textit{a priori} unknown) dual space. Indeed, this expression assumes that we are performing the new path integration over string space coordinates in a curved target space, suitably dual to pure Yang-Mills (or close enough to make statements thereupon, e.g. possessing a mass gap and confinement). We will explain how to choose one later on. It is worth mentioning at this stage that whatever space we choose, however the meson operator should be constructed in that case, the actual nature of the prefactor $W$ is of little importance, only the $\exp\left( {ik_i^\mu X_\mu(\sigma_i)}\right)$ part is truly relevant. Alone, this operator is tachyonic, and in flat space superstring theory it should be removed. But, other than removing a tachyonic pole in the amplitude, the function $W$ affects the kinematics of the amplitude very little. Indeed its effect is (in flat space, and so we must assume also in curved space) to soak up Lorentz-group quantities like polarisation vectors (absent here when we deal with scalars) and to affect the intercept of the Regge function, but strictly nothing more. We will therefore involve it very little in the computation to come, and assume, once the time comes to notice the analyticity properties of the amplitude, that this operator should be there to remove tachyonic poles.

The end result is a string-theoretic expression that involves nothing but the expectation value of an operator, inside this outer integral over all possible points of insertion. The fact that these operators are formulated in an explicitly position-space specific fashion is a potential problem for perturbation theory, but we will see how to work around it. From this point, we need to choose a specific string background to perform a detailed computation, but, given the nature of spaces that exhibit confinement physics, there are a few general statements about the rest of the procedure that we can explain without the need to choose a background. Let us sketch the procedure.

\section{A generic description of the procedure}

The broad lines of the corrective procedure, the set-up of a perturbative expansion on the worldsheet, can be formulated wtihout the need to specify a background. The properties that a given string background needs to exhibit in order for dual Wilson loops on the boundary of the space to have an area-scaling law are known explicitly \cite{Kinar:1998vq}. In detail: assume a translation-invariant Wilson loop along the time direction, label the other (finite) spatial direction $x$ and the holographic coordinate $U$. Then, the shape of the Wilson loop is prescribed by the function $U(x)$ (or $x(U)$, but metrics over such spaces typically are written in terms of $U$). Labelling $f^2(U(x)) = G_{00} G_{xx}$ and $g^2(U(x))=G_{00}G_{UU}$, the confinement criterion can be written as follows. Assuming without loss of generality that the end of physical space is at $0$:

\begin{theorem}
Let $f,g$ be smooth positive functions over $ 0< U < \infty$, and at $0$ assume the following form
\begin{equation}
f(U)=f(0) + U^k a_k + O(U^{k+1})\,\,\, , \,\,\, g(U)=U^j b_j + O(s^{j+1})
\end{equation}
with $f(0)>0$, $k>0$, $a_k>0$, $b_j>0$ and $j\geq-1$. Also assume that $\int \frac{g}{f^2}$ converges at infinity. \\

Then, $k\geq 2(j+1)$, implying an even geodesic line descending from $U=\infty$ exists and has energy proportional to its length, for large enough values of the length.
\end{theorem}

To avoid dealing with potentially pathological backgrounds, we assume that the spaces we will be dealing with will contribute functions $f,g$ that are also analytic at $0$, with one exception- a simple pole in $g$ can also be treated by a convenient change of variables, an example of which will be seen in the treatment of Witten's model. Fractional powers (or worse behaviour) in a series expansion around $0$ can lead to unstable solutions to the variational problem, which is an undesirable feature in an analysis of fluctuations around a classical solution. Higher-order poles are not allowed by the requirements of the theorem, without which no such even geodesics exist. There are therefore two main classes of case studies to handle with these assumptions (many discussions on the subjects usually refer to just those cases altogether):
\begin{itemize}
\item $g$ is regular at $0$: $j>0$ implying $k\geq2$, therefore $f$ has a minimum at the end of space, and its value at the minimum is positive
\item $g$ has a simple pole at $0$, implying $k\geq 1$. 
\end{itemize}
We will see an example of both cases in practice. Note that, if $f$ has a minimum, one generically expects $k=2,j=0$: unless the space has been constructed in such a way that lower order coefficients vanish, the functions $f,g$ have no reason to have a very high number of early zero coefficients in their expansion. Furthermore, the first case in which there is a pole in the metric, it is necessary in order for us to proceed to regularise the coordinate singularity, removing the pole. This is done in the treatment of Witten's model later on, and one can observe that once this is done the space obeys the conditions for the second case with a saturated bound for $k$. Thus, in general, $k=2\,\, ,\,\,\, j=0$ should be considered to be the most generic case of confining spaces.

There is however another obstacle in dealing with such spaces, notably, that the metric can have zero eigenvalues at the point of interest. Certainly, if the space presents a horizon singularity, the above prescription regularises it, but the Jacobian for the transformation vanishes at $0$. If not, generically the determinant of the metric happens to vanish at the origin from the beginning. Indeed, whichever coordinate system is used, such spaces are often generated from branes wrapping compact surfaces, causing cone-like submanifolds to exist in the overall geometry, Typically, this takes the form  of a warped cone where the base is a compact submanifold of the space (typically a sphere or an equivalent space), whose radius vanishes at the point of interest, where the string worldsheets accumulate, like so:

\begin{equation}
ds^2 = \cdots + U^2B(U) d\Omega^2 \,\,\, \implies\,\,\, \lim\limits_{U\rightarrow 0}\det(G)= 0
\end{equation}
for some regular, non-vanishing function $B$. This is problematic: to enforce reparametrisation invariance at the  quantum level, a non-linear sigma model's path integration measure is proportional to $\det (G)$:
\begin{equation}
\mathcal{Z}=\int \left[\prod_M DX^M\right]\exp(-S)\,\,\,\,\,\mbox{where}\,\,\, \left[\prod_M DX^M\right]=\sqrt{\det(g)}\prod_M DX^M
\end{equation}
 If this vanishes around the classical solution, the quantum theory is meaningless. One can interpret this determinant factor as an effective potential on the coordinate fields by exponentiating it and adding it to the action by 
\begin{equation}
\det(G)=\exp\left(\int\log(G) \right) 
\end{equation}
and this potential, diverging at 0, completely destabilises the vacuum.

  We therefore need to flatten or "unwrap" this cone, to rewrite the surface in terms of coordinates in which the metric is invertible at the point of interest, that is, in the would-be vacuum state of the non-linear sigma model. This can be systematically done,  as will be shown in two examples below, based on an approach similar to the Kruskal coordinate transformation in General Relativity. The idea is to perform the following change of coordinates, we start with a non-straight cone over (typically) an n-sphere, and re-express it in terms of flat Cartesian coordinates, up to the appearance of a radial conformal factor $C$, like so:\begin{equation}
ds^2=A(U)dUdU + U^2 B(U) dS_n^2 = C\left( \sum_{i=1}^{n+1} Z_i^2\right)  \left(\sum_i^{n+1} dZ_idZ_i \right)
\end{equation}
This will help, indeed, if we assume that the cone is locally straight around $0$, that is, $A,B$ are constant to leading order, then so must $C$ be constant at leading order, as a straight cone over a sphere is immediately diffeomorphic to flat space. Then this submanifold's newly expressed metric no longer has a vanishing determinant, removing the offending part of the global space's determinant.

The procedure to perform this transformation is well known and in principle applicable globally throughout the space, however, actually computing the new coordinates explicitly over the totality of the space can prove to be very difficult. On the other hand, we are interested in an effective action for a string worldsheet lying at the region of interest, for which the deviations away from this vacuum state are assumed to be parametrically small in order for perturbation theory to be valid. Therefore, we only need to obtain an approximate version of the metric around the relevant region, and perform the Kruskal-like operation we have detailed up to a certain order in our approximation. Whichever parameters we employ to materialise the relative smallness of the excitations then become coupling constants between the various fields in the metric.

Lastly, one does not expect the branes to generically wrap all of the compact directions, as such high-dimensional compact spaces have difficult geometries to handle. We therefore also expect that there are compact directions whose radii do not vanish. This does not upset the field theory, and the string worldsheet could probe these dimensions, but the relevant parts of the action will be difficult to deal with, and either way should give less relevant contributions to the path integral than the unbounded holographic coordinate. We are therefore not going to include them in our effective action, although in principle one should. The mindset should be to think that these spaces are usually artefacts of the construction of the string backgrounds, they do not typically represent desired symmetries of the dual theory.

To recapitulate, the steps required to convert the form of the metric into a nice perturbative effective Lagrangian for the worldsheet theory are the following:

\begin{enumerate}
\item Formulate the metric in such a way as to make apparent the irregularities of the metric, in particular taking care to show which submanifolds are expected to shrink to zero size.  
\item Around the end of space, expand the coefficients of the metric order by order in the size of the holographic directions, which will become a coupling constant.
\item Unwrap the cone: perform a change of variables to coordinates where there is no vanishing submanifold. This usually cannot be done globally but is simple to do in this expansion.
\item Canonically normalise the kinetic terms, express the interaction terms in the effective Lagrangian and write Feynman rules.
\end{enumerate}

Once this is done, we have all the ingredients for an effective field theory. The vacuum is well-defined in the partition function, and the action systematically looks like flat space plus some interaction terms left over from the metric, whose small coupling constant is directly related to the curvature of the space. We can proceed to insert our ansatz for the meson operators and start computing the relevant expectation value. Let us do so for an instructive first example, Witten's model.

\section{Results for Witten's backreacted D4 brane background}

We will now detail how the above prescription is applied in a simple case, Witten's background of $D4$ branes wrapping a thermal circle, as described in \cite{Witten:1998zw}. This case is a good candidate for a first investigation of this nature: its geometry is simple to describe, the end of space is characterised by an explicit length scale (the position of the black hole horizon), the compact coordinates describe a fairly simple submanifold, and importantly, it features a singularity, so it illustrates how one is able to deal with them.
\subsection{Description of the space, classical result}

The space has a simple structure, it has a four-dimensional Anti-deSitter type boundary, but the branes generating curvature also happen to wrap an $S_1$ subspace. The remaining 4 directions are compactified on an $S_4$ forming the base of a cone in the holographic direction. The metric for the space is the following:
\begin{equation}
ds^2=\left(\frac{U}{R}\right)^{3/2} \left(d X^2+\text{d$\tau $}^2 f(U)\right)+\left(\frac{R}{U}\right)^{3/2}\left( \frac{d U^2 }{f(U)}+ U^2d\Omega _4\right) \text{ , where }f(U)=1-\frac{U_{\text{KK}}^3}{U^3}
\end{equation}

As advertised previously, this is a valid candidate for our purpose as it exhibits a coordinate singularity at $U=U_{KK}$, satisfying the confinement condition. One can check that, to zero-th order, this does indeed reproduce the Veneziano amplitude, as was done in \cite{Armoni:2015nja}. We will summarise the result for clarity: we assume $U=U_{KK}, dU=0$ in a way that keeps the metric consistent, and ignore the influence of the extra 4-sphere, the result is a string action in flat space (albeit with a rescaled string tension):

\begin{equation}
S=\left(\frac{1}{2\pi \alpha^\prime}\right) \left(\frac{U_{KK}}{R}\right)^{3/2}\int d^2 \sigma \, \,  \partial_\alpha X^\mu  \partial^\alpha X_\mu
\end{equation}
And it is well-known that the flat open string 4-point function reproduces the Veneziano amplitude. Our task is now to improve that, and to include fluctuations of the worldsheet around this horizon.

\subsection{Preparing the space for a perturbative expansion}

First, we deal with the pole in the metric. It is a simple pole, therefore we change variables $U=U_{KK}(1+\frac{u^2}{U_{KK}^2})$, as previously explained, this ensures the metric is regular at that point.

\begin{align}
ds^2=&\left(\frac{U_{KK}}{R}\right)^{3/2}\left(1+\frac{u^2}{U_{KK}^2} \right)^{3/2}  \left(d X^2+\left(1-\frac{1}{\left(1+\frac{u^2}{U_{KK}^2} \right)^3 }\right) d\tau^2 \right)\label{varchange} \\ 
&+\left(\frac{R}{U_{KK}}\right)^{3/2}\frac{1}{\left(1+\frac{u^2}{U_{KK}^2}\right)^{3/2}}\left(4u^2d u^2 \left(1-\frac{1}{\left(1+\frac{u^2}{U_{KK}^2} \right)^3 }\right)^{-1} + U_{KK}^2 (1+\frac{u^2}{U_{KK}^2})d\Omega _4\right)\nonumber
\end{align}

We introduce $\lambda=\frac{U_{KK}}{R}$ for brevity, \textit{a priori} its magnitude is not fixed, though we will have to do so later. Let us check the regularity of the metric: after simplifications, we have
\begin{align}
G_{XX}&=\lambda ^{3/2}+\frac{3 \lambda ^{3/2} u^2}{2 U_{KK}^2}+O\left(\frac{u^4}{U_{KK}^4}\right)\nonumber\\
G_{\tau\tau}&= \lambda ^{3/2}\left(\frac{u^2}{U_{KK}^2}+1\right)^{3/2} \left(1-\frac{1}{\left(\frac{u^2}{U_{KK}^2}+1\right)^3}\right)=\frac{3 \lambda ^{3/2} u^2}{U_{KK}^2}\left( 1+\frac{u^2}{2 U_{KK}^2}\right)^{-1} +O\left(\frac{u^6}{U^6_{KK}}\right)\label{expansion}\\
G_{uu}&=\frac{4}{ \lambda ^{3/2}}\frac{\left(\frac{u^2}{U_{KK}^2}+1\right)^{3/2}}{\frac{u^4}{U_{KK}^4}+3 \frac{u^2}{U_{KK}^2}+3}=\frac{4}{3 \lambda ^{3/2}}\left( 1+\frac{ u^2}{2U_{KK}^2}\right) +O\left(\frac{u^4}{U^4_{KK}}\right).\nonumber
\end{align}

So we have dealt with the horizon but at the cost of making $G_{\tau\tau}$ vanishing with $u$, thus the determinant vanishes at zero, as explained this spoils the path integration measure. However, the shape of the $(u,\tau)$ submanifold is nice enough to be able to allow for the previously mentioned "unwrapping" of the subspace via Kruskal coordinates.

Let us introduce the following notation: $G(u)=\frac{u^2}{U_{KK}^2}\frac{(1+\frac{u^2}{U_{KK}^2})^{3/2}}{(1+\frac{u^2}{U_{KK}^2})^{3}-1}\mbox{ , }A=\frac{3\lambda^{3/2}}{U_{KK}^2}\mbox{ , }B=\frac{4}{3 \lambda ^{3/2}}$, such that the $(\tau,u)$ submanifold metric is of the form:

\begin{equation}
ds^2=Au^2\frac{d\tau^2}{G(u)}+BG(u)du^2.
\label{nearflat}
\end{equation}

The Kruskal procedure starts by defining a new radial variable, expressed in terms of a quantity usually called Tortoise coordinate, defined as the following integral
\begin{equation}
F(u)=\int_0^u \frac{1}{\nu}(G(\nu)-1)d\nu.
\end{equation}
Unfortunately, for the full expression of $G_{uu}$ described above, this integral is not readily expressible in closed form. Either way, as explained previously, only the local behaviour of the metric around the region $u=0$ is of relevance, so for $G$ we will substitute its power series expansion as detailed in Eq.\ref{expansion}, i.e. $G(u)=1+\frac{u^2}{U_{KK}^2}$. In which case, the Tortoise coordinate is straightforward to compute and
\begin{equation}
F(u)=\frac{u^2}{4U^2_{KK}}.
\end{equation}

We will not express $F$ in the following to keep with the generality of the procedure. Indeed, let us now implement the "unwrapping": we want to switch to pseudo-Cartesian coordinates parametrising a conformally-flat 2D space, let us introduce

\begin{equation}
Y+iZ=ue^{i\sqrt{\frac{A}{B}}\tau}\exp(F(u)),
\end{equation}
one can then check that the flat $(Y,Z)$ metric obeys
\begin{equation}
dY^2+dZ^2=\exp\left( 2F(u)\right) \left( G^2(u)du^2+\frac{A}{B}u^2d\tau^2\right) 
\end{equation}

After multiplication by the conformal factor $B\exp(-2F(u))\frac{1}{G(u)}$, this expression reproduces the original metric of Eq.\ref{nearflat}, up to the fact that the left hand side mixes old and new coordinates, for this we make use of the following relation

\begin{equation}
Y^2+Z^2=u^2\exp(2F(u))
\end{equation}
Which in theory can be inverted for a sensibly-defined $F(u)$. In our case the inversion gives

\begin{equation}
u=\sqrt{2}U_{KK}\sqrt{\mathcal{W}_L\left(\frac{Y^2+Z^2}{2U_{KK}^2} \right) }
\end{equation}

Where $\mathcal{W}_L$ is Lambert's W function, a.k.a the product log. Substituting for this expression, we arrive at a fully-formed change of variables:
\begin{align}
\frac{4}{3 \lambda ^{3/2}}\frac{\exp(-\mathcal{W}_L\left(\frac{Y^2+Z^2}{2U_{KK}^2} \right))}{1+\mathcal{W}_L\left(\frac{Y^2+Z^2}{2U_{KK}^2} \right)}\left(dY^2+dZ^2 \right)&= \frac{4}{3 \lambda ^{3/2}}\left( 1+\frac{ u^2}{2U_{KK}^2}\right)du^2+\frac{3 \lambda ^{3/2} u^2}{U_{KK}^2}\left( 1+\frac{u^2}{2 U_{KK}^2}\right)^{-1} d\tau^2\label{newmtric}
\end{align}

In writing it this way we make manifest the fact that the new metric is $U(1)$ invariant as it should be, since this was present in the original metric as a shift-invariance of the $\tau$ coordinate. We can therefore package up the new coordinates in a complex variable or $SO(2)$ doublet, which we will call $\Upsilon=Y+ i Z$. Now, the equality in Eq.\ref{newmtric} should be understood as only being true for small deviations around zero, i.e. for small $\frac{u}{U_{KK}}$ and $\frac{|\Upsilon|}{U_{KK}}$. We will therefore compute the new metric elements to first order in a series expansion in this approximation. We encapsulate the result in the new form of the metric, ignoring the ancillary compact coordinates:
\begin{align}
ds^2=\lambda^{3/2}(1+\frac{3|\Upsilon|^2}{2 U_{KK}^2})dXdX+\frac{4}{3\lambda^{3/2}}\left(1-\frac{|\Upsilon|^2}{U_{KK}^2} \right) d\bar{\Upsilon}d\Upsilon+\dots + O\left( \frac{|\Upsilon|^4}{U_{KK}^4}\right) 
\end{align}

We also check that the determinant is now non-vanishing at $u=0$:

\begin{equation}
\det(G)=U^8_{KK}\frac{16}{9\lambda^{3}}\left(1+6\frac{|\Upsilon|^2}{U^2_{KK}}\right) \det\left( S_4 \right) 
\end{equation}

This term can be added to the string worldsheet action: exponentiating its logarithm and Taylor-expanding it 
in our fashion, this generates a parametrically small mass term for the new radial field:

\begin{equation}
m^2=\frac{9\lambda^{3/2}}{2TU^2_{KK}}
\end{equation}
Note the appearance of an inverse factor of the length scale $U_{\text{KK}}$, this mass is therefore a very small number, this will be relevant later on. All in all, we will therefore be considering the following effective field theory:

\begin{align}
\mathcal{L}=&\lambda^{3/2}\partial_\alpha X^\mu \partial^\alpha X_\mu + \frac{4}{3\lambda^{3/2}}\partial_\alpha\bar{\Upsilon}\partial^\alpha\Upsilon + \frac{6}{T U^2_{KK}}|\Upsilon|^2\nonumber+\frac{3\lambda^{3/2}}{2U_{KK}^2}|\Upsilon|^2 \partial_\alpha X^\mu \partial^\alpha X_\mu-\frac{4|\Upsilon|^2}{3\lambda^{3/2}U^2_{KK}}\partial_\alpha\bar{\Upsilon}\partial^\alpha\Upsilon
\end{align}

There are a few comments that we should make before moving on to the computation, relating to supersymmetry. We have not discussed its influence on the procedure at all so far, and we cannot proceed unless we do.

\subsection{A word on Supersymmetry}

Despite the fact that Witten's background has explicit supersymmetry breaking, this is only applicable to the space-time theory, not the worldsheet sigma model. We should in all fairness be considering the following \textit{a priori} supersymmetric Lagrangian (where $\Psi$ is the superpartner of $X$)

\begin{align}
\mathcal{L}=\left(G_{AB}\partial X^A\partial X^B+ G_{AB}\bar{\Psi}^A\slashed{\partial} \Psi^B \right.  \left.+ \Gamma_{ACB}\bar{\Psi}^A\left(  \slashed{\partial}X^C\right)   \Psi^B + R_{ABCD}\left( \bar{\Psi}^A\Psi^C\right) \left( \bar{\Psi}^B\Psi^D\right) \right)
\end{align}

and then observe that it is eventually broken. Specifically, we should focus on the third term of the Lagrangian $\Gamma_{ACB}\bar{\Psi}^A\left(  \slashed{\partial}X^C\right)\Psi^C$: we are inserting purely bosonic $X^\mu$ operators in the path integral, so the most relevant interaction terms for us are those that couple those variables to the curvature. However, if we write the Christoffel symbol connecting $X^\mu$ to other variables we find

\begin{equation}
\Gamma_{AXB}=\left( \partial_B G_{AX} + \partial_X G_{AB} - \partial_A G_{BX}\right) 
\end{equation}

The middle term vanishes as $G$ depends only on the radial coordinate. We are left with a tensor that is antisymmetric in $A,B$, multiplying the quantity $\bar{\Psi}^A\gamma^\mu \Psi^B$, a symmetric object. This component therefore cancels, there is no direct coupling of the $X$ coordinates to the fermionic degrees of freedom, they will only start happening at high loop order via a mediating $\Upsilon$ excitation.

Note that this was proven using only the fact that the metric is diagonal and that it depends only on the holographic radial direction. These properties will still be true for a wide spectrum of potential backgrounds this process applies to and the comments made above still apply to them.

\subsection{Worldsheet Feynman rules and diagrammatic expansion}

We now turn to constructing a set of Feynman rules for our theory. We have a Lagrangian, but there is a crucial ingredient missing between it and the path integral at hand, the operator insertion. The exact expression to compute is the following:
\begin{align}
\mathcal{A}\left(k_{i} \right) =\oint \prod\limits_{i=1}^4d\sigma_i \int DX \,\, W(\sigma_i)\exp\left( {ik_i^\mu X_\mu(\sigma_i)}\right)  \exp\left( -\frac{1}{ 2\pi\alpha^\prime} {\int d^2\sigma\, G_{MN} \partial_\alpha X^M \partial ^\alpha X^N}\right).
\end{align}

As explained previously, the main purpose of the function $W$ (in general a function of the coordinates $X$ and their superpartners $\Psi$) is to affect the intercept of the resulting Beta function, and soaking up spin-related quantities such as polarisations and spinor components. This can be verified in flat space for any of the known operators of the open string spectrum, and since our operator is scalar and massless in our case there is no point in trying to guess its shape or its effect on the effective field theory- we will simply keep in mind that the tachyon pole of the amplitude is removed. It cannot affect the functional form of the Beta function, it cannot even affect the slope of the Regge trajectory. We therefore feel justified in not keeping track of it very much, and consider the effect of the insertion of $\exp\left( {ik_i^\mu X_\mu(\sigma_i)}\right)$.

In flat space these operator insertions are usually dealt with by turning them into a non-zero current for the $X^\mu$ fields and performing some Gaussian eliminations to obtain the Veneziano amplitude. If we want to preserve this term as the "classical" result, in order to find perturbations around it, we need to take care to reproduce many of the same steps, notably, dealing with non-zero currents. The way we do it is by writing the operators in this way:

\begin{equation}
\exp\left( {i\sum\limits_{j=1}^4 k_j^\mu X_\mu(\sigma_j)}\right) = \exp\left(\int d^2\sigma \, J_0(\sigma)^\mu X_\mu (\sigma) \right) \, ,\,\, J_0^\mu(\sigma)= \sum\limits_{j=1}^4 i k_j^\mu \delta(\sigma-\sigma_j)
\end{equation}

 The goal will then be to compute the partition function of the theory, evaluated at a non-zero value for the $X$ current. In particular we really only desire those diagrams involving particles sourced by this current and not the vacuum bubbles that usually pop up in the computation, so we should really divide the sourced partition by the partition function at zero current. Then, we set up the usual interaction picture with the help of currents. We split the expression for the \textit{total} current for $X$ as $J=J_0 +\delta J$: $J_0$ is the constant, classical value of the current and $\delta J$ represents small variations around the classical value, this allows us to implement quantum fluctuations around this non-vacuum classical ground state of the string.
 
We move to normalised units for the various fields by setting $\alpha=1$ and rescaling every coordinate by the value of the metric at $u=0$ to have canonical normalisation. Then, schematically, the following happens, encapsulating all the interaction terms in a potential $V(X)$:
\begin{align}
&\mathcal{Z}[J]=\int [DX]\exp\left( -\int\left({1\over 2\pi} X\Delta X \right) +\int\left(J_0\cdot X+\delta J\cdot X \right) -{1\over 2\pi} \int V(X)\right)  \nonumber\\
&=\int [DX]\exp\left( -{1\over 2\pi}\int V(\frac{\delta}{\delta J})\right) \exp\left( -\frac{1}{2\pi}\int \left(X-\sqrt{\frac{\pi}{\Delta}}(J_0+\delta J) \right) \left(X-\sqrt{\frac{\pi}{\Delta}} (J_0+\delta J)\right) \right.\nonumber\\
&\left.-\frac{\pi}{2}\int \delta J \Delta^{-1}\delta J-\frac{\pi}{2}\int J_0\Delta^{-1}J_0-\int \delta J\cdot J_0\right)\nonumber\\
&=\int \exp\left( -\frac{\pi}{2} J_0\Delta^{-1}J_0\right) \exp\left( \frac{\pi}{2}\int V(\frac{\delta}{\delta J})\right)\exp\left( -\frac{\pi}{2}\int \delta J\Delta^{-1}\delta J-\int \pi \delta J\Delta^{-1}J_0\right) 
\end{align}
The above is then evaluated at $\delta J=0$ i.e. $J = J_0$, the small $\delta J$ expansion creates the usual Feynman series of terms interpreted diagrammatically. This machinery has the advantage of isolating a leading factor $\int \exp\left( -\frac{\pi}{2}  J_0\Delta^{-1} J_0\right)$, the classical result, yielding the usual Veneziano amplitude when the expression for $ J_0$ is substituted in and the surface integration is performed. However, this adds a twist to the construction of Feynman rules, due to the term $\delta  J\Delta^{-1} J_0$: while its symmetric companion implements propagation of an $X$ particle from one vertex to another, this term immediately stops a propagator exiting a vertex and inserts the contents of the current. This term implements particles being spontaneously created and destroyed out of the underlying non-zero current, effectively they become a type of 1-leg vertex. Since we are computing the partition function, there are no ingoing or outgoing states that reach infinity, so every "external" leg will need one of these current insertions to stop it. We are still technically computing "vacuum" diagrams, except it is a vacuum of quantum fluctuations defined around a non-zero classical current.

In detail, here is the rule-set we will be using to construct diagrams.

\begin{itemize}
\item \includegraphics[width=0.15\textwidth]{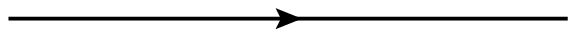} : $X^\mu$ propagator, $\delta^{\mu\nu} \dfrac{1}{p^2}$
\item \includegraphics[width=0.15\textwidth]{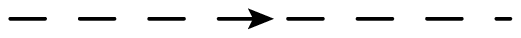} : the $\Upsilon$ propagator, $\delta^{ij}\dfrac{1}{q^2+m^2}$
\item \includegraphics[width=0.06\textwidth]{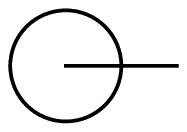} : the 1-leg  $ J_0^\mu(\sigma)$-insertion vertex, $\lambda^{-3/4}\sum\limits_{i=1}^4 k_i^\mu\exp(ip\cdot\sigma_i)$
\item \includegraphics[width=0.1\textwidth]{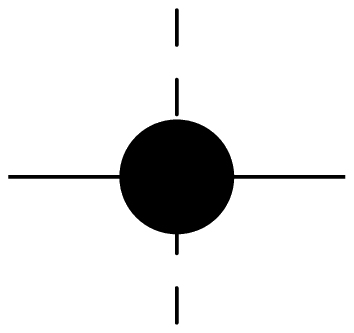} : the 4-leg $X$-interaction vertex: $\delta^{\mu\nu}\delta^{ij}\dfrac{9\lambda^{3/2}}{8U_{KK}^2}(p_i\cdot p_j)$ 
\end{itemize}

With these rules we can start computing contribution up to any order. However, this scenario has its share of idiosyncrasies, notably that loop order should not be trusted to control the relative order of the diagrams. Given the momenta associated to the interaction vertex, it is easy to prove that every diagram we compute must be logarithmically divergent, the degree of divergence is not controlled by loop order here. This is clear from dimensional analysis through the fact that these diagrams are being generated by the partition function, a dimensionless object. Only our expansions parameters control the size of the diagrams and therefore we can order them differently depending on our choices. By inspection of the Feynman rules it appears that the scaling of a diagram with $V$ vertices and $2E$ "outer" legs is

\begin{equation}
\left( \lambda\right) ^{3\left(V-E \right)/2 } \left(\frac{1}{U_{\text{KK}}^2} \right)^{V} =  \left(\frac{\lambda^{3/2}}{U_{\text{KK}}^2}\right)^V \left(\lambda^{-3/2} \right) ^{E}
\end{equation}

\begin{figure}[h]
\centering
\includegraphics[width=0.50\textwidth]{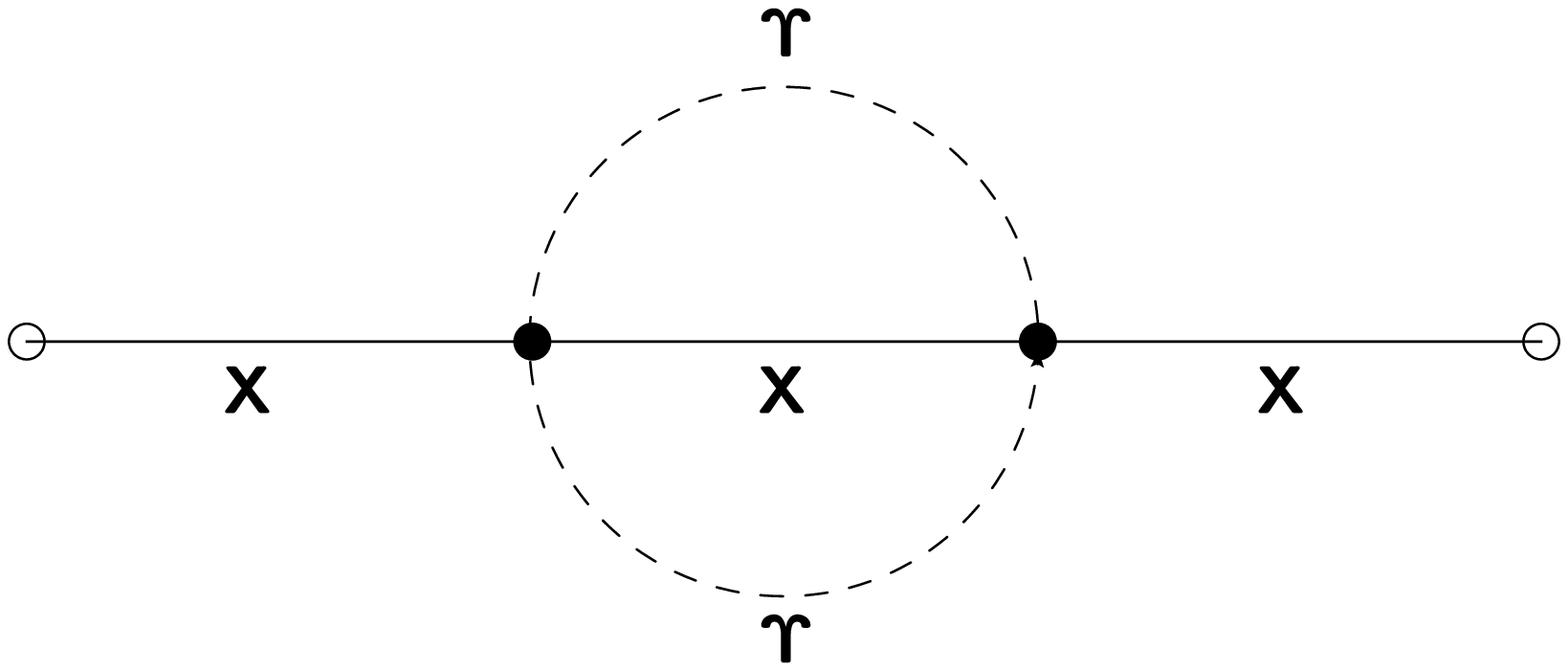}
\includegraphics[width=0.30\textwidth]{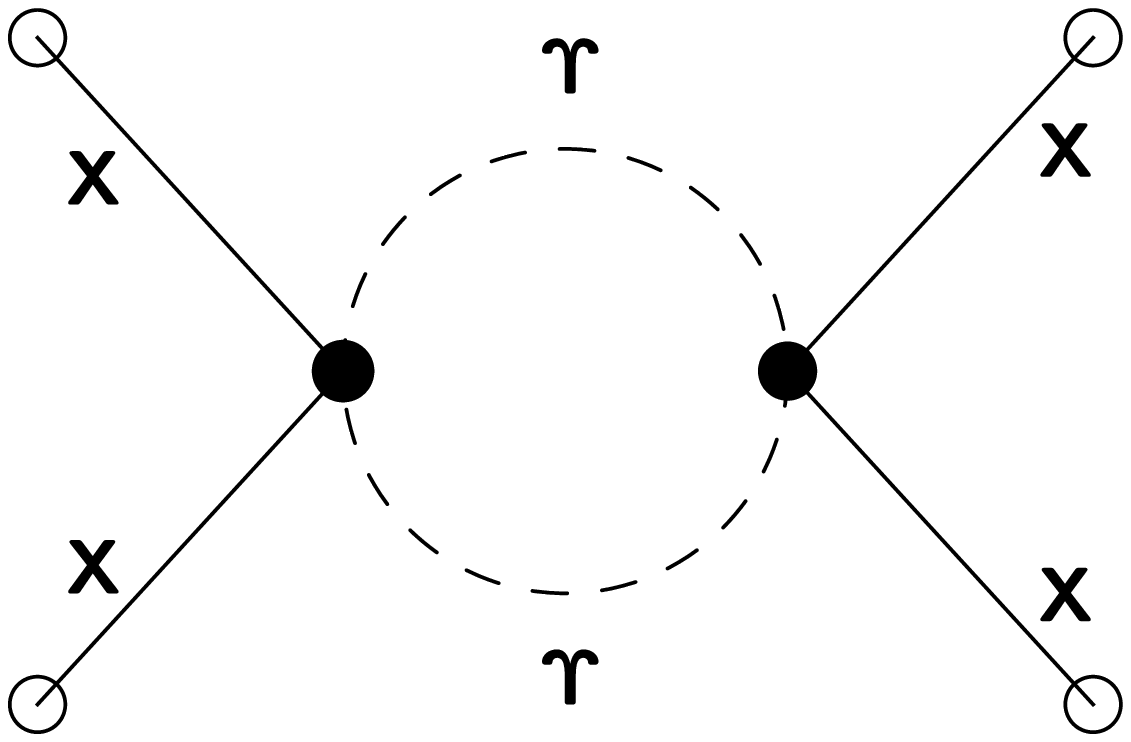}
\caption{A two-loop diagram of lower order than a one-loop diagram}
\label{loopdiags}
\end{figure}

The scaling was rewritten in such a way as to isolate the dependence on $V$ and $E$, thus, a diagram's order can be made to be classified by those two numbers very directly. The more of each it has, the higher order it is, as long as $\lambda^{3/2}\gg U_{\text{KK}}^2\gg 1$. For instance, it is possible to make certain one-loop diagrams higher order than certain two-loop diagrams, an example is given in Fig.\ref{loopdiags}

The first diagram is manifestly two-loop, but has half as many "outer" legs as the second, thus they contribute respectively at order $\left(\frac{\lambda^{3/2}}{U_{\text{KK}}^2}\right)^2 \left(\lambda^{-3/2} \right) ^{1}$ and $\left(\frac{\lambda^{3/2}}{U_{\text{KK}}^2}\right)^2 \left(\lambda^{-3/2} \right) ^{2}$.

Now, it is not clear whether the second, one-loop diagram is actually non-zero or even well-defined in our framework, but in the case that it does give a finite answer, this scaling behaviour allows us to ignore it. This is helpful in a sense- we ought to be computing corrections to the 2-point function, as that is the correlation that yields (at tree level) the Veneziano amplitude. Either way, we choose to focus on those diagrams rather than ones with more external legs, whatever the loop order.

\begin{figure}[h]
\centering
\includegraphics[width=0.4\textwidth]{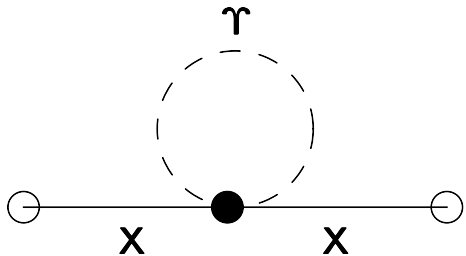}
\caption{A diagram with no momentum transfer}
\label{straightdiag}
\end{figure}

On the other hand, diagrams with too low loop order turn out to be too simple to contribute anything qualitatively new. Indeed, a broad class of diagrams involving no momentum transfer between the virtual particles and "external" states, such as the one showed in Fig.\ref{straightdiag}, only contribute quantitative, not qualitative, corrections to the amplitude.

This means that the momentum integral factorises neatly in a bubble- or tadpole-like contribution involving the loop particles only, times a simple tree-like contribution involving the "external" particles. To justify why these diagrams are less interesting than others, we need to delve into the details of the regularisation and subtraction scheme we will employ, as we will not be doing standard dimensional regularisation.

\subsection{Analytic regularisation of 2D integrals and the trivial nature of factorisable diagrams}

From dimensional analysis it is clear to see that the diagrams we will express are all logarithmically divergent. These types of divergences are a fairly common feature in 2d quantum field theories, starting with the propagator. For reasons explained above, worldsheet 2-point corrections are much lower in our expansion order than higher-point diagrams, so we can hope to deal with all the logarithmic divergences we will encounter in one go with a good choice for a regularisation scheme, namely one that gives the correct, finite value for the massless propagator with no subtractions required. If even the propagator is divergent and requires a counter term in the action, any other diagram with N propagators will then have at most $2^N$ variants where one includes or not the counterterm for each line, which greatly increases the amount of effort required to complete any computation, let alone prepare it for a systematic order-by-order tower of corrections.

For this purpose we will employ, instead of the standard dimensional regularisation, another form of dealing with divergent integrals, which is analytic regularisation. The prescription is the following, we replace the standard momentum-space propagator by its analytically regulated version:

\begin{equation}
\frac{1}{p^2}\rightarrow \lim\limits_{x\rightarrow 0}\dfrac{d}{dx}\left( x \mu^{-x}\frac{1}{(p^2)^{1-x}} \right)
\end{equation}

This for each propagator, introducing many such parameters $x_1,x_2\dots$ all being sent to zero. We have introduced $\mu$, a mass scale added to soak up the extra dimensions, identically to dimensional regularisation. This is technically slightly stronger than a regularisation scheme, it does more than expose the divergence in the propagator, it actively removes it, without handling any counter-terms. We will, throughout the remaining, use $D_{x=0} \hat{=} \frac{d}{dx}{\big\rvert}_{x=0}$.

With this definition, let us compute said propagator and check that this does give a finite value:

\begin{align}
\Delta(\sigma)&=D_{x=0} x \mu^{-x} \int \frac{d^2p}{(2\pi)^2} \frac{e^{ip\cdot\sigma}}{(p^2)^{1-x}}=D_{x=0} x \mu^{-x} \int d \alpha \frac{\alpha^{-x}}{\Gamma(1-x)}\int \frac{d^2p}{(2\pi)^2} \exp\left( ip\cdot \sigma-\alpha p^2\right)  \nonumber\\
&=\frac{1}{2\pi}D_{x=0} x \mu^{-x} \int d \alpha \frac{\alpha^{-x-1}}{\Gamma(1-x)} \exp \left( -\frac{\sigma^2}{4\alpha} \right) = \frac{1}{2\pi}D_{x=0} x \mu^{-x} \int du \frac{u^{x-1}}{\Gamma(1-x)}\exp\left(- \frac{\sigma^2}{4}u \right) \nonumber\\
&= \frac{1}{2\pi}D_{x=0} x \mu^{-x}\frac{\Gamma(x)}{\Gamma(1-x)} \left(\frac{4}{\sigma^2} \right)^{x}=\frac{1}{2\pi}D_{x=0} \frac{\Gamma(1+x)}{\Gamma(1-x)} \left(\frac{4}{\mu \sigma^2} \right) ^{x}\nonumber\\
&= \frac{1}{2\pi} \left( -\gamma_E +\log(4) -\log(\mu) - \log(\sigma^2)\right) 
\end{align}

As is often done we can rescale $\mu$ (chosen arbitrarily) to absorb the constant factors, as per the $\overline{\text{MS}}$ procedure. We will do this and more: in order to simplify later computations we will also dimensionally rescale $\mu$: the total operation is

\begin{equation}
\mu \rightarrow \mu 4e^{-2\gamma_E} m^2
\end{equation}

Where $m$ is the mass of the $Y,Z$ particles. The propagator becomes
\begin{equation}
\Delta(\sigma)=-\frac{1}{2\pi}\log(m^2 \sigma^2)
\end{equation}
once the arbitrary scale is removed. With this choice, the mass appears to influence the massless propagator: this is a feature, while $\sigma$ is dimensionless in the units that we chose, it can be rescaled to have dimensions of length and so this choice of scaling makes sense. In addition, this will help in computing the more trivial types of diagrams. Particularly, we look at the simplest diagram we can create with these Feynman rules, the 1-loop 2-point graph shown in Fig.\ref{straightdiag}. Removing the overall constants and outer Fourier kernel, the central part of the integratin subsumes to computing

\begin{equation}
\int \frac{d^2p_1 d^2 p_2 d^2q}{(2\pi)^6}\frac{(p_1\cdot p_2)\delta(p_1-p_2)}{p_1^2 p_2^2 (q^2+m^2)}
\end{equation}

As advertised this splits neatly into the $p_i$ integration, which, after applying the $\delta$ function, is trivially just another copy of the propagator, and a bubble-type integral over $q$. With our choice of regularisation and subtraction, the result is
\begin{equation}
\frac{1}{(2\pi)^2}\int \frac{d^2q}{q^2+m^2}\rightarrow \frac{1}{4\pi}\left(\log(4)-3\gamma_E \right) 
\end{equation}

Again, this seems odd as this massive bubble contribution has no dependence on the mass, this is due to the version of $\overline{\text{MS}}$ we are using here.

All in all, this diagram merely contributes a finite-value contribution to the wavefunction renormalisation, i.e. a finite shift of the string tension $T$. But, either way, the bare value of the string tension is in theory fixed \textit{a posteriori} to match real-world parameters, so these shifts are of much lesser interest to us. They still occur, and for the remainder of the exercise we will consider $T$ to be no longer the bare value but its effective value after all such simple shifts allowed by our constraints. There is no modification of the effective Veneziano amplitude at the level of these diagrams, let us proceed further to those that do feature momentum transfer and less trivial integrals to compute.

\subsection{A diagram with momentum transfer}

\begin{figure}[h!]
\centering
\includegraphics[width=0.650\textwidth]{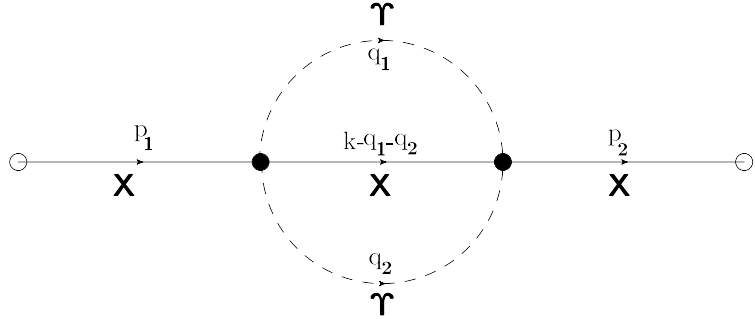}
\caption{The first diagram with momentum transfer and a qualitative correction}
\label{firstcorrec}
\end{figure}

Now we return to the first diagram of Fig.\ref{loopdiags}, shown again in detail in Fig.\ref{firstcorrec}. It does possess momentum transfer, therefore will not factorise trivially, thus has potential to generate some new behaviour in the amplitude we seek.

This corresponds to the following integral:
\begin{align}
I= \frac{81\lambda^{3/2}}{32 U_{KK}^4}\left( \sum\limits_{i<j}k_i\cdot k_j\right) &\int \frac{d^2 p_1 d^2 p_2 d^2 k d^2 q_1 d^2 q_2}{(2\pi)^{10}}e^{i(p_1\cdot\sigma_i - p_2 \cdot \sigma_j )}\nonumber\\
&\times \frac{ \delta(p_1-k) \delta(k-p_2) (p_1\cdot (k-q_1-q_2))(p_2\cdot(k-q_1-q_2))}{p_1^2 p_2^2 (q_1^2+m^2) (q_2^2+m^2) (k-q_1-q_2)^2}\nonumber\\
=\frac{81\lambda^{3/2}}{32 U_{KK}^4}\left( \sum\limits_{i<j}k_i\cdot k_j\right) &\int \frac{d^2p d^2q_1 d^2 q_2}{(2\pi)^6} \dfrac{e^{i(p\cdot \Delta)}(p\cdot(p-q_1-q_2))(p\cdot(p-q_1-q_2))}{p^4(q_1^2+m^2) (q_2^2+m^2) (p-q_1-q_2)^2}
\end{align}
where $\Delta=(\sigma_i - \sigma_j)$.

Disregarding temporarily the outer Fourier transform over $p$, the central element of this integral corresponds to a class of diagrams usually called "Sunset" or "London Transport" diagram, here in the case where two out of the three inner legs are massive and one is massless. At best, this is a rather non-trivial integral. Certainly, in 4D dimensional regularisation, a $(d-4)$ expansion has been obtained, as shown for instance in \cite{Smirnov:2012gma}. This technique however involves a high amount of extra machinery to obtain, besides, we are in 2D and are using analytic regularisation. Switching regularisation schemes would cost us having to deal with a propagator counter-term in order to define it with a finite value, thus drastically increasing the amount of computation required.

On the other hand, the \textit{massless} diagram is particularly easier to compute. Indeed, the mass in question, as we have explained, is parametrically small, of the same order as the interaction strength between the two types of particles. It is in fact almost an inconsistency of our approximations to have considered a fully-resummed propagator, arguably we should have treated the mass as a 2-point vertex and left it that way for consistency. This poses further conceptual difficulties in computing the "bubble" contribution of the above, so we will refrain from doing so until this point. Now, we will consider that the mass is indeed very small and can (to first order) be ignored. Doing so here does not break the theory, momentum transfer in the diagram ensures that the new singularities created in this process do not escape our regularisation scheme's scope. It is possible that in doing so, we create IR divergences of equal and opposite contribution to the leading UV divergence, thus artificially raising the power of the overall divergence, as has been pointed out to us. This does not change the following results qualitatively.

Therefore, we will consider this simplified integral:

\begin{equation}
\hat{I}=\frac{81\lambda^{3/2}}{32 U_{KK}^4}\left( \sum\limits_{i<j}k_i\cdot k_j\right) \int \frac{d^2p d^2q_1 d^2 q_2}{(2\pi)^6} \dfrac{e^{i(p\cdot \Delta)}(p\cdot(p-q_1-q_2))(p\cdot(p-q_1-q_2))}{p^4q_1^2 q_2^2 (p-q_1-q_2)^2}
\end{equation}

The numerator structure can be simplified significantly by using the Laporta algorithm, implemented in Mathematica via the popular \texttt{FIRE.m} package. After reduction, we obtain

\begin{align}
\hat{I}=&\frac{81\lambda^{3/2}}{32 U_{KK}^4}\left( \sum\limits_{i<j}k_i\cdot k_j\right) \int \frac{d^2p d^2q_1 d^2 q_2}{(2\pi)^6} \frac{p^4}{6}\dfrac{e^{i(p\cdot \Delta)}}{p^4q_1^2 q_2^2 (p-q_1-q_2)^2}\nonumber\\
=&\frac{27\lambda^{3/2}}{64 U_{KK}^4}\left( \sum\limits_{i<j}k_i\cdot k_j\right) \int \frac{d^2p d^2q_1 d^2 q_2}{(2\pi)^6} \dfrac{e^{i(p\cdot \Delta)}}{q_1^2 q_2^2 (p-q_1-q_2)^2}
\end{align}

The $p$ integral is as usual a final Fourier transform to obtain a position-space result, focusing on the $(q_1,q_2)$ subsector, we apply the regularisation scheme and compute the integral at arbitrary analytic coefficients. Using Schwinger's $\alpha$ parametrisation:

\begin{align}
I_{\text{sub}}&\hat{=}\int \frac{ d^2q_1 d^2 q_2 }{(2\pi)^4}\dfrac{1}{(q_1^2)^{1-x}(q_2^2)^{1-y}((p-q_1-q_2)^2)^{1-z}}\\
&=\int  \frac{ d^2q_1 d^2 q_2 }{(2\pi)^4} \int_{(0,0,0)}^{(\infty,\infty,\infty)} d\alpha d\beta d\gamma \exp(-\alpha q_1^2 -\beta q_2^2 -\gamma(p-q_1-q_2))\frac{\alpha^{-x}\beta^{-y}\gamma^{-z}}{\Gamma(1-x)\Gamma(1-y)\Gamma(1-z)}\nonumber
\end{align}

The quadratic form in $q_{1,2}$ has determinant $\alpha \beta + \beta \gamma +\gamma \alpha$, eliminating it leaves over a $p$-dependent term $\frac{\alpha \beta \gamma p^2}{\alpha \beta + \beta \gamma +\gamma \alpha}$ such that we get after momentum integration
\begin{align}
I_{\text{sub}}=\frac{1}{(2\pi)^2\Gamma(1-x)\Gamma(1-y)\Gamma(1-z)}\int_{(0,0,0)}^{(\infty,\infty,\infty)} d\alpha d\beta d\gamma \frac{\alpha^{-x}\beta^{-y}\gamma^{-z}}{\left( \alpha \beta + \beta \gamma +\gamma \alpha\right) }\exp\left( -\frac{\alpha \beta \gamma p^2}{\alpha \beta + \beta \gamma +\gamma \alpha}\right) 
\end{align}
It is conventional at this point to introduce a new integrand $T$ equal to the sum of any subset of $\alpha,\beta,\gamma$ like it is done in many such formulae involving Gamma and Beta functions. It is a property of such integrals that it does not matter which subset of the variables we choose to sum up, this feature is sometimes called the Cheng-Wu theorem, physically it is understood as a manifestation of world-line parametrisation invariance. We then rescale every variable by $T$, implicitly done in the following:

\begin{align}
I_{\text{sub}}&=\int \frac{d\alpha d\beta d\gamma dT}{(2\pi)^2\Gamma(1-x)\Gamma(1-y)\Gamma(1-z)} \delta(1-(\dots)) \frac{\alpha^{-x}\beta^{-y}\gamma^{-z}}{\left( \alpha \beta + \beta \gamma +\gamma \alpha\right) } T^{2-x-y-z}e^{ -\left(\frac{\alpha \beta \gamma p^2}{\alpha \beta + \beta \gamma +\gamma \alpha}\right) }\nonumber\\
&= \frac{1}{(p^2)^{1-x-y-z}}\frac{\Gamma(1-x-y-z)}{(2\pi)^2\Gamma(1-x)\Gamma(1-y)\Gamma(1-z)}\int d\alpha d\beta d\gamma \delta(1-(\dots)) \frac{\alpha^{y+z-1}\beta^{x+z-1}\gamma^{x+y-1}}{\left( \alpha \beta + \beta \gamma +\gamma \alpha\right)^{x+y+z}}\nonumber\\
&=  \frac{1}{(p^2)^{1-x-y-z}}\frac{\Gamma(1-x-y-z)\Gamma(x)\Gamma(y)\Gamma(z)}{(2\pi)^2\Gamma(1-x)\Gamma(1-y)\Gamma(1-z)\Gamma(x+y+z)}
\end{align}

The above identity of Gamma functions can be checked with Mathematica or by hand, it is helpful in that case to choose $T=\alpha+\beta$ or any two (but not three or one) variables, which allows for separation of the integral into two Euler Beta functions, through which there are some simplifications of numerators and denominators.

In the next step, we now perform the Fourier transform of the above function, as can be seen this is close to the standard propagator, but the regulator structure around it will prevent it from being trivial.

\begin{align}
\tilde{I}&\hat{=}\frac{1}{(2\pi)^2}\mu^{-x-y-z}\int \frac{d^2 p}{(2\pi)^2} \frac{\Gamma(1-x-y-z)\Gamma(1+x)\Gamma(1+y)\Gamma(1+z)}{\Gamma(1-x)\Gamma(1-y)\Gamma(1-z)\Gamma(x+y+z)}\frac{e^{ip\cdot\Delta}}{(p^2)^{1-x-y-z}}\nonumber\\
&=\frac{\Gamma(1+x)\Gamma(1+y)\Gamma(1+z)}{(2\pi)^2\Gamma(1-x)\Gamma(1-y)\Gamma(1-z)\Gamma(x+y+z)}\mu^{-x-y-z}\int \frac{d^2 p}{(2\pi)^2} \int d \alpha \exp\left(ip\cdot\Delta -\alpha p^2 \right) \alpha^{-x-y-z}\nonumber\\
&=  \frac{\Gamma(1+x)\Gamma(1+y)\Gamma(1+z)}{(2\pi)^3\Gamma(1-x)\Gamma(1-y)\Gamma(1-z)\Gamma(x+y+z)}\mu^{-x-y-z}\int d \alpha \alpha^{-1-x-y-z} \exp\left( -\frac{\Delta^2}{4\alpha}\right) \nonumber\\
&= \frac{\Gamma(1+x)\Gamma(1+y)\Gamma(1+z)}{(2\pi)^3\Gamma(1-x)\Gamma(1-y)\Gamma(1-z)\Gamma(x+y+z)}\mu^{-x-y-z}\int d u u^{x+y+z-1} \exp\left( -\frac{\Delta^2 u}{4}\right) \nonumber\\
&=  \frac{\Gamma(1+x)\Gamma(1+y)\Gamma(1+z)}{(2\pi)^3\Gamma(1-x)\Gamma(1-y)\Gamma(1-z)} \left( \frac{4}{\mu\Delta^2}\right)^{x+y+z}
\end{align}
Where we have used $u=\frac{1}{\alpha}$ to complete the integral. At this stage the value of the diagram is given by the limiting operation described above, which essentially reduces to obtaining the $O(xyz)$ coefficient of the above analytic function as a triple power series. One would obtain, formally,
\begin{equation}
\tilde{I}= \frac{1}{(2\pi)^3}\left( -\gamma_E + \log\left(\frac{4}{\mu \Delta^2} \right) \right)^3 
\end{equation}

With the rescaling in action, and restoring the constants from the Feynman rules, we finally get
\begin{equation}
I= -\frac{1}{(2\pi)^3}\frac{27\lambda^{3/2}}{64 U_{KK}^4}\left( \sum\limits_{i<j}k_i\cdot k_j\right)  \left(\log\left(\Delta^2 m^2\right)  \right)^3 
\end{equation}

We will package the numerical constants and expansion parameters for brevity's sake by defining

\begin{equation}
\rho^2 = \frac{1}{4\pi^2} \frac{27\lambda^{3/2}}{64 U_{KK}^4}.
\end{equation}

\subsection{Consequences for the amplitude}

Now that we have computed this quantum correction, we need to insert it in the outer path integral and sum over all insertion positions. The total amplitude will look like the following:

\begin{equation}
\mathcal{A}(k_i)=\delta\left(\sum_i k_i \right) \oint \prod_{i=1}^{4} d \sigma_i \exp\left(-\frac{1}{2}J_0\Delta^{-1}J_0+ \dots \right) 
\end{equation}
Where $(\dots)$ imply further corrections. The leading order computation proceeds by exponentiating the log term given by the propagator, reducing the integral to a Euler Beta-type integral and performing it. Higher powers of the log will not be conveniently treated by this approach so we use an expansion of this overall integral to write our computation as 

\begin{equation}
\mathcal{A}(k_i)=\oint \prod_{i=1}^{4} d \sigma_i \exp\left(-\frac{1}{2}J_0\Delta^{-1}J_0\right)\left( 1 + \dots  \right) 
\end{equation}

Now, in our case there is a small difference. By our definition of the renormalisation scale, a mass-like constant appears in every logarithm. This will not prove to be difficult to manage, thankfully, it will affect every term in much the same way, providing an overall multiplicative constant through rescaling of the integral. For the leading term for instance:

\begin{align}
\mathcal{A}_0(k_i)&\hat{=}\oint \prod_{i=1}^{4} d \sigma_i \exp\left(-\frac{\pi}{2}J_0\Delta^{-1}J_0\right)\nonumber\\
&=\delta^{(4)}\left( \sum k_i \right) \oint \prod_{i=1}^{4} d \sigma_i \exp\left( -\sum_{i<j} k_i\cdot k_j \log\left( m^2|\sigma_i-\sigma_j|^2\right) \right)\nonumber\\
\end{align}

The momentum-conserving delta-function appears in the second line due to the zero modes of $X^\mu$. We are then free to choose the positions $\sigma_i$ in this integral: we will take 

\begin{equation}
\sigma_1= z\text{ , }\sigma_2=0\text{ , }\sigma_3=\frac{1}{m}, \sigma_4=\infty
\end{equation}

This will give one element of the total amplitude as it will involve only two out of the three Mandelstam variables, the total answer should sum over the permutations of the above choices. We now reduce to a one-dimensional integral:
\begin{align}
\mathcal{A}_0(k_i)\hat{=}\int_0^{1/m} dz  \left| mz\right|^{2 k_1\cdot k_2} \left(m \left|\frac{1}{m}-z\right|\right)^{2 k_1\cdot k_3} +\left( \text{other combinations of }k_i\right)  \\
\end{align}

We here introduce the standard notation of Mandelstam variables. Considering every momentum as ingoing, we write
\begin{equation}
s= (k_1+k_2)^2 = 2k_1\cdot k_2\,\,,\,\,t=(k_1+k_3)^2=2 k_1\cdot k_3\,\, , \,\, u=(k_1+k_4)^2=2k_1\cdot k_4
\end{equation}

We now only have to define $w= m z$, turning the integration variable dimensionless, to recover a properly Beta-type integral. $\mathcal{A}_0$ changes by a single factor of $m$, but since we are computing the value of the partition function, overall constants can always be removed by normalisation. From the above one then recovers the usual Veneziano amplitude:

\begin{equation}
\mathcal{A}_0(s,t,u) = \dfrac{\Gamma(s)\Gamma(t)}{\Gamma(s+t)} + (\text{perms.})
\label{ampint}
\end{equation}

Here we have used the fact that we have not inserted a tachyonic operator. The function $W$, previously ignored, should ensure that this is the correct result- this has no tachyonic pole, the first pole is a massless one, as required.

Now, we can investigate the nature of the first correction, given the prescription described in all of the above.
\begin{align}
\mathcal{A}_1(s,t)= \int_0^1 dz \left|z\right|^{s-1}\left|1-z\right|^{ t-1}\left(s\log^3\left(|z|^2 \right) +t\log^3\left( \left|1-z\right|^2\right)    \right) 
\end{align}

These two integrals subsume to different versions of the following identity

\begin{equation}
\int_{0}^{1} dw w^{x-1}(1-w)^{y-1} \log^n(x)= \partial^n_x B(x,y)
\end{equation}

So that the result becomes

\begin{equation}
\mathcal{A}_1(s,t)= \left(1- \rho ^2 \left( s\partial_s^3 + t \partial_t^3\right)  \right) B(s,t)
\end{equation}

In general, multiple derivatives of the Beta function can be expressed through the use of the family of functions $\psi^{(n)}$ defined by $\psi^{(0)}=\Gamma^\prime / \Gamma \,\,\, , \psi^{(n+1)}=\psi^{(n)\prime}$. This gives in our case

\begin{align}
B^{(3,0)}(x,y)= &B(x,y)\left((\psi ^{(0)}(x)-\psi ^{(0)}(x+y))^3-\psi ^{(2)}(x+y)+\psi ^{(2)}(x)\right.\nonumber\\
&\left.+3 (\psi ^{(1)}(x)-\psi ^{(1)}(x+y))
   (\psi ^{(0)}(x)-\psi ^{(0)}(x+y))\right)  
\end{align}

We would then like to observe the asymptotics of this corrected amplitude, in order to try and find a new form for the Regge trajectory. For this purpose we write the large $s$, fixed $t$ asymptotic expansion of the above expression, often called the Regge limit of the amplitude. Indeed, one expects that the 4-point amplitude in this limit has the following behaviour

\begin{equation}
\mathcal{A}^{(4)} \sim\alpha(s)^{\alpha(-t)}
\end{equation}

We therefore expand the result in this limit and match the expressions to obtain a new, modified form of $\alpha(s)$. We recall the asymptotic behaviour of the $\psi^{(n)}$ functions:

\begin{equation}
\psi^{(0)}(z) \sim \log(z)\,\,\,,\,\,\, \psi^{(n>0)}(z)\sim z^{-n}
\end{equation}

Other than $\psi^{(0)}$ all the other functions vanish at infinity, so we expect that the former produce the main contribution. It happens that the functions $\psi^{(0)}(-s)$ are generated by the $t-$derivatives in the expressions above. We then get:

\begin{equation}
	\mathcal{A}(s,t)\sim s^t-\rho^2 t \log(s)^3 \sim s^{t(1-\rho^2 \log(s)^2)}
\end{equation}

It is interesting to note at this point that this is exactly the result one obtains by taking the $t-$derivatives of the approximate form of the Beta function in the Regge limit. This shows that the Regge limit and our corrective operations commute. This will be relevant in a following section on higher point scattering amplitudes, for which no explicit analytical expressions are known and the corrective operations must be taken after the Regge limit.

This suggests that the Regge function is modified in the following fashion
\begin{equation}
	\alpha(s)= s^{1-\rho^2 \log^2(s)}
\end{equation}

\begin{figure}[h!]
	\centering
\includegraphics[width=0.5\textwidth]{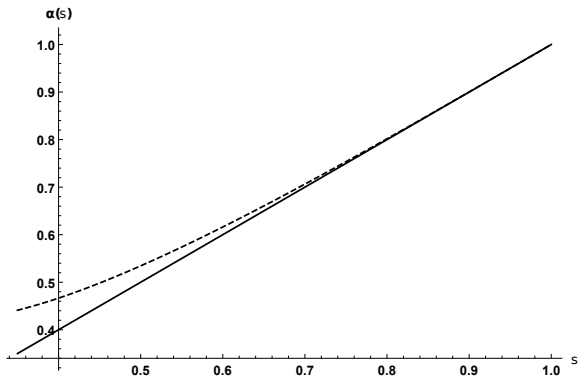}
\caption{The bending effect of stringy corrections, shown for $\rho^2=0.2$}
\label{bending}
\end{figure}

 In this form the behaviour of the Regge trajectory can be plotted and is shown in Fig.\ref{bending}.
These types of behaviours are commonly seen in experimental results of QCD, furthermore, they were also seen previous studies \cite{Imoto:2010ef,Sonnenschein:2016pim}. Note that in those examples, $M^2$ was being plotted against $J$, given that $J=\alpha(M^2)$ this is the other way around compared to our plot here, but all the curves bend slightly towards the same axis ($J$ or $\alpha(s)$). 

This concludes the analysis of this particular model: we have found the expected generic interaction term and analysed its influence, which looks consistent with the physics of actual mesons. For consistency, it is interesting to try another explicit example of the procedure, one with a less obvious vanishing submanifold structure and an explicit end of space (no horizon). The Klebanov-Strassler background is a good choice, we will focus on it.

\section{The case of the Klebanov-Strassler background}

Now that we have explained the generic procedure with a specific example, we can continue to iterate these steps on other backgrounds that feature confining physics. Over the past decade there has been a great deal of interest in the Klebanov-Strassler string background \cite{Klebanov:2000hb}, we will apply our technique here.

\subsection{Description of the space, separation of coordinates}

Compared to the previous case the space features some more non-trivial geometry. The space is based around the 5-dimensional fibre bundle manifold called resolved conifold, topologically equivalent to a cone whose base is $S_3\times S_2$, but where the size of the $S_3$ does not vanish at the origin of the holographic coordinate. Its metric is canonically written as follows:
\begin{align}
ds^2&= h^{-1/2}(\tau)dx^\mu dx_\mu + h^{1/2}(\tau) ds^2_6 \,\,\, \text{, where}\nonumber\\ h(\tau)&=2^{2/3}(g_s M\alpha^\prime)^2 \epsilon^{-8/3}\int_{\tau}^{\infty} dx \left( \frac{x \coth x -1}{\sinh^2 x}\left(\sinh (2x) -2x\right)^{2/3}\right)
\label{KSBG}
\end{align}
where $ds^2_6$ is the metric of this deformed conifold space, 
\begin{align}
ds^2_6 =& \frac{\epsilon^{4/3}}{2}K(\tau)\left( \frac{1}{3K^3(\tau)}\left( d\tau^2+(g^5)^2\right)+ \cosh^2\left(\frac{\tau}{2}\right)\left( (g^3)^2 + (g^4)^2\right) \right.\nonumber\\
&\left.+ \sinh^2\left(\frac{\tau}{2} \right) \left((g^1)^2 +(g^2)^2 \right)   \right) \,\,\, , \,\,\,  K(\tau)=\frac{(\sinh(2\tau)-2\tau)^{1/3}}{2^{1/3}\sinh\tau}
\label{conifold}
\end{align}
where the $g^{i=1\dots5}$ are (quite verbose) differential one-forms, such that $(g^1)^2 +(g^2)^2$ is topologically a two-sphere and $(g^3)^2 + (g^4)^2 +(g^5)^2$ a three-sphere. Explicitly,

\begin{align}
g_1=& \frac{1}{\sqrt{2}}\left( d \theta _2 \sin (\psi )-d \phi _2 \sin \left(\theta _2\right) \cos (\psi
   )-d \phi _1 \sin \left(\theta _1\right)\right)  \nonumber\\
g_2=& \frac{1}{\sqrt{2}}\left( -d \theta _2 \cos (\psi )-d \phi_2 \sin \left(\theta _2\right)
   \sin (\psi )+d \theta _1 \right)  \nonumber\\
g_3=& \frac{1}{\sqrt{2}}\left( -d \theta _2 \sin (\psi )+d \phi _2 \sin \left(\theta _2\right) \cos (\psi
   )-d \phi _1 \sin \left(\theta _1\right)\right) \\
g_4=& \frac{1}{\sqrt{2}}\left( d \theta _2 \cos (\psi )+d \phi _2 \sin \left(\theta _1\right) \sin (\psi
   )+d \theta _1 \right)  \nonumber\\
g_5&= d \phi _1 \cos \left(\theta _1\right)+d \phi _2 \cos \left(\theta _2\right)+d
   \psi \nonumber\\
\end{align}

From the asymptotics of the various functions, this space reproduces the advertised behaviour close to the radial origin. This is, however, a little unsuitable for our purposes, two steps are required in order to bring it to a form that is adequate. Firstly, it is not directly obvious which limit needs to be taken for the space to flatten as in the previous case, since the radial coordinate is naturally dimensionless. This will be made clearer at a later stage of the process.

Secondly, the metric of this space is vanishing at the origin, $\det(g)\sim \sinh^4 \tau$, due to the presence of the vanishing two-sphere. While topologically the one-forms $g^{1,2}$ do describe the sphere in question, we will require to unwrap this cone into a three-dimensional asymptotically flat space, which requires that the two angles parametrising the sphere be decoupled "well enough" from the other coordinates. That is, they should be written so that we can see a separation of the space into a set of two coordinates describing only motion in the $S_2$, and three for motion in the $S_3$, up to fibration of the former over the latter. In that way, we can fix the three angles of the non-vanishing sphere to a constant value (i.e. we ignore motion in this direction, the base space of the fibre), and have an explicitly cone-like metric that will transform nicely under the Kruskal procedure. The canonical parametrisation, derived from the original construction of the conifold space as a metric over the coset space $\frac{SU(2)\times SU(2)}{U(1)}$ does not reproduce this behaviour very clearly.

This reparametrisation of the deformed conifold metric exists, thankfully, and the space was described in terms of very adequate coordinates in the literature \cite{Krishnan:2008gx}. The details are as follows, we introduce the $S_2$ specific coordinates $(\theta,\phi)$ and those for the $S_3$ as $(\alpha,\beta,\gamma)$. The deformed conifold metric described in Eq.\ref{conifold} can be entirely rewritten as the following:
\begin{align}
ds^2_6=& \epsilon^{4/3}K(\tau)\left( \frac{1}{6K^3(\tau)}\left( d\tau^2+(h^3)^2\right)+ \frac{\cosh^2\left(\frac{\tau}{2}\right)}{4}\left( (h_1)^2 + (h_2)^2\right) \right.\nonumber\\
&\left.+ \sinh^2\left(\frac{\tau}{2} \right) \left(\left(d\theta-\frac{1}{2}h_2 \right)^2 +\left(\sin\theta d\phi -\frac{1}{2} h_1 \right)^2    \right)\right) 
\end{align}
where, crucially, the differential forms $h_{i=1\dots3}$ are spanned by $d\alpha,d\beta,d\gamma$. It is therefore obvious in this formalism that if we forbid motion in those directions the remaining angular variables form an explicitly parametrised $S_2$, whose radius vanishes with $\tau$. From this starting point it will then be straightforward to apply the Kruskal procedure.

\subsection{Near-flat limit of the metric and Kruskal change of coordinates}

Due to the complicated nature of the metric coefficients, in order to compute the correct change of coordinates we will once again need to expand the metric around $\tau=0$. The crucially difficult part in doing so is evaluating the integral present in $h(\tau)$. No closed form is known for it, but the integrand vanishes exponentially, therefore the zeroth-order term can be easily computed numerically. The first order term vanishes by construction (implementing the confinement condition) and the second order term can be computed manually by repeated differentiation. The form of the expansion is known in the literature, \cite{Basu:2012ae}, to wit

\begin{equation}
h(\tau)= 2^{2/3} \alpha^2 \epsilon^{-8/3}(a_0-a_2\tau^2) + O(\tau^3)\,\,\, , \,\,\, \alpha=g_s M \alpha^\prime\,\,\, , \,\,\, a_0\simeq 0.71805\,\,\, , \,\,\, a_2=2^{-1/3}3^{-4/3}
\end{equation}
 Further coefficients can in theory be computed by taking higher derivatives but in principle we would like to stop here. The function $K$ is analytic and so easily expandable. 

To put the $X$ and $\tau$ variables on the same footing we will rescale both of them such that the metric elements approach unity at $\tau=0$, implicitly done in all of the following. In terms of these rescaled variables, the metric elements are as follows:
\begin{align}
G_{xx}&= 1+\frac{\sqrt[3]{6} a_2}{  a_0^{3/2}}  \frac{\tau^2}{\alpha}+ O\left( \tau^4\right) \nonumber\\
G_{\tau\tau}&= 1 +\frac{\sqrt[3]{6} \left(2 a_0-5 a_2\right)}{5a_0^{3/2}} \frac{\tau^4}{\alpha}+ O\left( \tau^4\right) \\
G_{S_2}&=  \tau^2\left( 1+ \frac{\left(a_0+30 a_2\right) \tau^2}{5\ 6^{2/3} \alpha  a_0^{3/2}}\right) + O\left( \tau^6 \right) \nonumber
\end{align}

Clearly we see that the leading order terms in these series define a flat cone in terms of the radial and angular variables, thus the leading order contribution to the amplitude will again be the flat open string scattering result.

Defining $A^2(\tau)= G_{\tau\tau}\,\, , \,\, B^2(\tau)= \frac{1}{\tau^2} G_{S_2}$ we are tasked with unwrapping a metric of the following functional form

\begin{equation}
ds^2= A^2(\tau)d\tau^2 + \tau^2 B^2(\tau) \left(d^2\theta + \sin^2 \theta d\phi^2  \right) 
\end{equation}

For this purpose we define
\begin{align}
C(\tau)&=\int_{0}^{\tau} \, \frac{1}{u} \left(\frac{A(u)}{B(u)}-1 \right)du \nonumber \\
U&= \tau \exp\left( C(\tau)\right) \sin \theta \cos \phi \nonumber\\
V&= \tau \exp\left( C(\tau)\right) \sin \theta \sin \phi \nonumber\\
W&= \tau \exp\left( C(\tau)\right) \cos \theta 
\end{align}
and write the flat metric in terms of the Kruskal coordinates:

\begin{align}
dU^2 + dV^2 + dW^2& = \exp(2C(\tau))\left( d\tau^2 \left(1 + \tau \times\frac{1}{\tau}\left( \frac{A(\tau)}{B(\tau)}-1\right) \right)^2 + \tau^2 \left(d\theta^2 + \sin^2\theta d\phi ^2 \right)   \right)\nonumber\\
&= \exp(2C(\tau))\left( d\tau^2 \left( \frac{A(\tau)}{B(\tau)} \right)^2 + \tau^2 \left(d\theta^2 + \sin^2\theta d\phi ^2 \right)   \right)
\label{ksmetricchange}
\end{align} 
from which we can multiply by a conformal factor $\exp(-2C(\tau))B^2(\tau)$ to obtain our starting metric on the right-hand side. As before the left-hand side still mixes the new and old coordinates and we must therefore make use of the radial relation

\begin{equation}
U^2 + V^2 + W^2 = \tau^2 \exp\left( 2 C(\tau) \right)
\end{equation}
which can be inverted in order to substitute for $\tau$ in the left-hand side above. As a short-hand and to make global symmetries explicit we will write the Kruskal coordinates, now promoted to quantum fields, as an $SO(3)$ vector $\Upsilon=(U,V,W)$. In terms of the expressions for the functions at hand we have the following relations:

\begin{align}
A(\tau)&=\sqrt{ 1 +\frac{\sqrt[3]{6} \left(2 a_0-5 a_2\right)}{5a_0^{3/2}} \frac{\tau^2}{\alpha} }+O(\alpha^{-2})=1 +\frac{\sqrt[3]{6} \left(2 a_0-5 a_2\right)}{10a_0^{3/2}} \frac{\tau^2}{\alpha} +O(\alpha^{-2}),\\
 B(\tau)&=\sqrt{1+ \frac{\left(a_0+30 a_2\right) }{5\ 6^{2/3}  a_0^{3/2} }\frac{\tau^2}{\alpha} }+O(\alpha^{-2})=1+ \frac{\left(a_0+30 a_2\right) }{10\ 6^{2/3}  a_0^{3/2}} \frac{\tau^2}{\alpha} + O(\alpha^{-2}) ,\\
C(\tau)&=\frac{\left(11 a_0-60 a_2\right) }{2 \left(a_0+30 a_2\right)}\log \left(\frac{\left(a_0+30 a_2\right) \tau ^2}{10\ 6^{2/3} \alpha a_0^{3/2}}+1\right), \\
\tau^2&= |\Upsilon| ^2\left( 1+\frac{|\Upsilon| ^2}{\alpha} \left(\frac{\sqrt[3]{6} a_2}{  a_0^{3/2}}-\frac{11}{10\ 6^{2/3}   \sqrt{a_0}}\right)\right) +O(\alpha^{-2})
\end{align}

So that the final expression for the asymptotically flat metric is the following:

\begin{equation}
ds^2= \left(1+\frac{\sqrt[3]{6} a_2}{ \alpha a_0^{3/2}} |\Upsilon|^2 \right)dX\cdot dX + \left(1+ \left(\frac{9  a_2}{6^{2/3}a_0^{3/2}}-\frac{6^{-2/3}}{\sqrt{a_0}}\right) \frac{|\Upsilon| ^2}{\alpha}\right) d\Upsilon\cdot d\Upsilon +\cdots 
\end{equation}
implying that we suppress higher order terms and non-vanishing compact directions.

In addition, we compute the approximate form for the determinant, obtaining a small mass for the $\Upsilon$ field: ignoring leading order constants that can be removed by normalisation of the path integral, 

\begin{equation}
\det(G)=\left(1+\frac{\sqrt[3]{3} \left(a_0+11 a_2\right) \Upsilon ^2}{2^{2/3} \alpha  a_0^{3/2}} \right) 
\end{equation}

It is sufficient to observe that the determinant has the correct form and that the mass will be parametrically small for the arguments of the previous section to continue to hold.

\subsection{Brief description of the field theory}

The effective Lagrangian that we will study is formally the same up to the numerical coefficients, and the normalisation of the fields. As explained previously, since this space saturates the bound between the power law indices of the metric coefficients (the generic case for such spaces) this will inevitably lead to an interaction term that has the shape we determined in the previous section. The Feynman rules will therefore be completely equivalent up to numerical constants. Interestingly one finds two parameters which play exactly the same roles as the $\lambda$ and $U_{\text{KK}}$ parameter we found previously in Witten's model.

Here is a recap of the Lagrangian:
\begin{align}
\mathcal{L}_{KS}= \partial_\alpha X^\mu \partial^\alpha X_\mu + \partial_\alpha \Upsilon^i \partial^\alpha \Upsilon_i + m^2 \Upsilon_i \Upsilon^i+ \frac{6^{1/3}a_2}{a_0^{3/2}}\frac{|\Upsilon|^2}{\alpha} \partial_\alpha X^\mu \partial^\alpha X_\mu
\end{align}

This leads to analogous Feynman rules, to wit

\begin{itemize}
\item \includegraphics[width=0.15\textwidth]{rules3.eps} : $X^\mu$ propagator, $\frac{1}{p^2}$
\item \includegraphics[width=0.15\textwidth]{rules4.eps} : the $\Upsilon$ propagator, $\frac{1}{q^2+m_\Upsilon^2}$
\item \includegraphics[width=0.06\textwidth]{rules1.eps} : the 1-leg  $J_0^\mu(\sigma)$-insertion vertex, $\sum\limits_{i=1}^4k^\mu_i \exp(p\cdot\sigma_i)\sqrt{\frac{\sqrt[3]{2} \alpha  \sqrt{a_0} }{\epsilon^{4/3}}}$
\item \includegraphics[width=0.1\textwidth]{rules2.eps} : the 4-leg $X$-interaction vertex: $\frac{\sqrt[3]{6} a_2}{ \alpha a_0^{3/2}} \left(  p_i\cdot p_j\right) $  ($p_i$ are momenta of $X$)
\end{itemize}

Now, observing the scaling of the various Feynman building blocks with $\alpha$ and $\epsilon$ one finds that a generic diagram has a scaling of 

\begin{equation}
\left(\frac{1}{\alpha} \right)^{V - J} \left(\frac{1}{\epsilon^{4/3}} \right)^{J} 
\end{equation}

This is much the same scenario as previously, with $\alpha$ taking the role of $U^2_{\text{KK}}$ and $\epsilon^{4/3}$ the role of $\lambda^{3/2}$. The same classification argument applies here too and we are allowed to remove higher-point functions from the computation at hand.

\section{Influence on higher-point functions}
\subsection{Defining the higher-point Veneziano amplitudes}
It is worth checking that the exact same effect is seen in the (multi-)Regge regime of higher-point functions. These additional n-point functions are defined by particular integral formulae based on the same very general analytic considerations that originally led to the discovery of the Veneziano amplitude \cite{Bardakci:1969cs}\cite{Chan:1969ex}. A modern, comprehensive review of the construction has been written \cite{DiVecchia:2007vd}, it features more general, less explicit expressions both for the stringy amplitude (in which $SL_2(\mathbb{C})$ invariance is made clear) and the heuristic expression (written elegantly as a very general integral on which we impose conditions based on Reggeon physics). We will not review all of it here and it will be sufficient to quote known results and properties from the already existing literature.

String theory predicts a scattering amplitude for open strings in flat space to have the following form

\begin{equation}
B_N = \int_{0}^1 \dots  \int_{0}^1\left( \prod\limits_{i=1}^N \theta\left(z_i-z_{i+1} \right) dz_i\right)  \left(\prod\limits_{i=2}^N\prod\limits_{j=i+1}^N \left( z_i - z_j\right)^{2p_i\cdot p_j}  \right) 
\end{equation}

Where $\theta$ is the step-function, imposing operator ordering between the various integration parameters. This formula is impractical, no known analytical expression exists to equate it for $N>4$, so to derive a Regge limit out of it is difficult. It needs to be brought to a more workable form. This is conveniently provided by the following, equivalent integral:

\begin{equation}
\int_{0}^1\cdots\int_{0}^1 \left( \prod\limits_{i=2}^{N-2} du_i u_i^{s_{1,i}-1}\left(1-u_i\right)^{ -s_{i,i+1}-1}  \right)\left(  \prod\limits_{i=2}^{N-3}\prod\limits_{j=i+2}^{N-1} \left(1-x_{ij} \right)^{s_{i,j}}\right) 
\end{equation}
where $x_{ij}=u_i u_{i+1} \dots u_j$ and $s_{i,j}=2p_i\cdot p_j$ are the various Mandelstam parameters in question. Although this form is suggestive of string scattering, it is derived from very general considerations of pole structure and Reggeon physics, some work is required to obtain  from it the open string scattering result with N operator insertions in flat space. The formulae are equivalent, as proved by Koba-Nielsen when formulating their decidedly more stringy integral \cite{Koba:1969kh}. This new integral provides a better setting to perform the Regge limit.

Indeed with this notation, the aforementioned limit of the amplitude is to take the following operation:
\begin{align}
s_{i,i+1}\gg 1\,\,\, \,\,\, s_{1i}=\text{const.} \,\,\, , \,\,\, i=2\dots N-2 \nonumber \\
\frac{s_{i,i+1}s_{i+1,i+2}}{s_{i,i+2}}=-k_i=\text{const.}
\end{align}.

We will sketch the arguments used to derive the behaviour of the integral. The key is to perform a change of variables of the form $u_i = 1- \exp\left(-\frac{v_i}{s_{i,i+1}} \right)$ and expanding to first order in the large Mandelstam variables, this affect each type of factor in the integrand in the following way:

\begin{align}
 u_i^{s_{1i}-1} &\rightarrow \left(s_{i,i+1} \right)^{-s_{1,i}+1}\times \left( v_i \right) ^{s_{1,i}-1} \nonumber\\
 (1-u_i)^{-e s_{i,i+1}}&\rightarrow \exp\left( - v_i\right) \exp\left( v_i/s_{i,i+1}\right)  \nonumber\\
\left(1-u_i\dots u_j \right)^{- s_{i,j}}&\rightarrow\left( 1- \frac{v_i\dots v_j}{s_{i,i+1}\dots s_{j,j+1}}\right)^{-\left(  k_i \dots k_j\right) \left(s_{i,i+1}\dots s_{j,j+1} \right)}\\
&\sim \exp\left( -k_iv_i\dots k_j v_j\right) \nonumber\\
du_i&\rightarrow \left( s_{i,i+1}\right)^{-1} \exp\left( -v_j/s_{i,i+1}  \right) \nonumber
\end{align}

The last terms coming from the measure cancel nicely some unwanted features of the first two lines, thus, extracting the lead Regge terms, the remainder terms do not depend on the asymptotically large Mandelstam variables at all, allowing us to write the amplitude as
\begin{equation}
\left( \prod\limits_{i=2}^{N-2}\left(  s_{i,i+1}\right) ^{-s_{1,i}}\right) \times G\label{multiregge}
\end{equation}
where $G$ is an integral depending only on the constant parameters defined in the rules of the limit. 

\subsection{Stringy corrections}

As was mentioned previously, there is no analytic solution to the integral at hand, hence the above integral machinery to express the Regge limit convincingly. But, it is sufficient to apply our correction term to this Reggeised form of the amplitude, indeed, the correction operation and the Regge limit commute, this was checked with the 4-point function, there is no reason to expect this to cease to be the case in higher point amplitudes. Agreeably, here too can the logarithmic terms appearing in the worldsheet action be pulled out of the integral by converting them to derivatives, thus we get (keeping in line with the previous notation):

\begin{equation}
\left(1- \rho ^2 \sum\limits_{i,j} s_{i,j}\left(\dfrac{\partial^3}{\partial s_{i,j} ^3} \right)  \right)B_N({s}) 
\end{equation}

This can be done even before the required manipulations to bring the string theory expression (the Koba-Nielsen integral) to the analytically favourable form described above. We are then free to do so and to then take the multi-Regge limit. Those derivatives that will contribute the most to the amplitude will be those depending on $s_{1,i}$ i.e. those variables we hold constant, as was the case previously. From the very factorised form of the multi-Regge limit above we get that for all $i=2\dots N-2$ each of these leading terms in Eq.\ref{multiregge} is modified to
\begin{equation}
\left( s_{i,i+1}\right) ^{\left(-s_{1,i}\left(1-\rho^2 \log^2\left(s_{i,i+1}\right)  \right) \right) }
\end{equation}
for every $i$. This is the exact same form that suggested the corrected Regge function in the 4-point case, across all Regge channels of the higher point amplitude, this is encouraging, we are recovering consistently the same effect across multiple sources.

\section{Conclusions}

In this paper we considered a computation of the 4-meson scattering amplitude using holography. It is motivated by a purely field-theoretical observation, namely the statement that assuming an area law for Wilson loops of all shapes and sizes produces a Veneziano amplitude, a condition that we can reproduce in holography but moreover find a way to systematically improve upon.

The prescription that we developed involves a calculation in curved space string theory. Since calculating string amplitudes in curved space is a very hard task, we used a perturbative expansion around a near-flat classical solution, and found the leading correction to the Veneziano amplitude. This allows us to extract a correction to the Regge function. We remark that this result is formally identical in many different string backgrounds, for reasons related to the nature of confinement physics and how they materialise in string theory. Finally we also observed that the result can be reobtained in any of the the multi- meson amplitudes. This is, on the whole, encouraging, motivating that the qualitative nature of the correction we found is a physical effect and not an artefact of the holographic duality.

We focused on the Regge regime, where flat space string theory is successful. The perturbative correction that we found is a qualitative improvement of the behaviour of the function in a comparatively lower energy regime. Our result agrees with previous results obtained by other approaches \cite{Imoto:2010ef,Sonnenschein:2016pim} which attempted to find a similar correction using different techniques.

We would also like to compare our approach with the more traditional ``flavour brane'' approach to quenched QCD. In the limit of large $N_c$ and and fixed $N_f$ the most popular approach to holographic QCD is to place a flavour brane in the background. Calculating meson scattering using this approach is a very difficult task, since it would involve a calculation in curved background with no obvious small parameter. Other difficulties include the coupling to RR flux and formulating vertex operators in curved space. Using our approach of a summation over string worldsheets we managed to circumvent all these problems. 

Encouraged by our result for meson-meson scattering we intend to address the problems of meson-baryon and baryon-baryon scattering.

\vspace{0.5cm}

\textbf{Acknowledgements.}$\,\,$ It is our pleasure to thank Timothy Hollowood and Carlos N\'{u}\~{n}ez for fruitful discussions. We would also like to thank the anonymous referee.

\vspace{0.5cm}

\newpage
\bibliographystyle{JHEP}
\bibliography{v5}

\providecommand{\href}[2]{#2}\begingroup\raggedright\begin{thebibliography}{10}

\bibitem{Veneziano:1968yb}
G.~Veneziano, {\it {Construction of a crossing - symmetric, Regge behaved
  amplitude for linearly rising trajectories}},  {\em Nuovo Cim.} {\bf A57}
  (1968) 190--197.

\bibitem{Maldacena:1997re}
J.~M. Maldacena, {\it {The Large N limit of superconformal field theories and
  supergravity}},  {\em Int. J. Theor. Phys.} {\bf 38} (1999) 1113--1133,
  [\href{http://arxiv.org/abs/hep-th/9711200}{{\tt hep-th/9711200}}]. [Adv.
  Theor. Math. Phys.2,231(1998)].

\bibitem{Makeenko:2009rf}
Y.~Makeenko and P.~Olesen, {\it {Wilson Loops and QCD/String Scattering
  Amplitudes}},  {\em Phys. Rev.} {\bf D80} (2009) 026002,
  [\href{http://arxiv.org/abs/0903.4114}{{\tt arXiv:0903.4114}}].

\bibitem{Armoni:2015nja}
A.~Armoni, {\it {Large-N QCD and the Veneziano Amplitude}},  {\em Phys. Lett.}
  {\bf B756} (2016) 328--331, [\href{http://arxiv.org/abs/1509.03077}{{\tt
  arXiv:1509.03077}}].

\bibitem{Armoni:2016nzm}
A.~Armoni and E.~Ireson, {\it {Holographic Corrections to the Veneziano
  Amplitude}},  \href{http://arxiv.org/abs/1607.04422}{{\tt arXiv:1607.04422}}.

\bibitem{Bigazzi:2004ze}
F.~Bigazzi, A.~L. Cotrone, L.~Martucci, and L.~A. Pando~Zayas, {\it {Wilson
  loop, Regge trajectory and hadron masses in a Yang-Mills theory from
  semiclassical strings}},  {\em Phys. Rev.} {\bf D71} (2005) 066002,
  [\href{http://arxiv.org/abs/hep-th/0409205}{{\tt hep-th/0409205}}].

\bibitem{Andreev:2004sy}
O.~Andreev and W.~Siegel, {\it {Quantized tension: Stringy amplitudes with
  Regge poles and parton behavior}},  {\em Phys. Rev.} {\bf D71} (2005) 086001,
  [\href{http://arxiv.org/abs/hep-th/0410131}{{\tt hep-th/0410131}}].

\bibitem{Polchinski:2001tt}
J.~Polchinski and M.~J. Strassler, {\it {Hard scattering and gauge / string
  duality}},  {\em Phys. Rev. Lett.} {\bf 88} (2002) 031601,
  [\href{http://arxiv.org/abs/hep-th/0109174}{{\tt hep-th/0109174}}].

\bibitem{Kinar:1998vq}
Y.~Kinar, E.~Schreiber, and J.~Sonnenschein, {\it {Q anti-Q potential from
  strings in curved space-time: Classical results}},  {\em Nucl. Phys.} {\bf
  B566} (2000) 103--125, [\href{http://arxiv.org/abs/hep-th/9811192}{{\tt
  hep-th/9811192}}].

\bibitem{Witten:1998zw}
E.~Witten, {\it {Anti-de Sitter space, thermal phase transition, and
  confinement in gauge theories}},  {\em Adv. Theor. Math. Phys.} {\bf 2}
  (1998) 505--532, [\href{http://arxiv.org/abs/hep-th/9803131}{{\tt
  hep-th/9803131}}].

\bibitem{Smirnov:2012gma}
V.~A. Smirnov, {\it {Analytic tools for Feynman integrals}},  {\em Springer
  Tracts Mod. Phys.} {\bf 250} (2012) 1--296.

\bibitem{Imoto:2010ef}
T.~Imoto, T.~Sakai, and S.~Sugimoto, {\it {Mesons as Open Strings in a
  Holographic Dual of QCD}},  {\em Prog. Theor. Phys.} {\bf 124} (2010)
  263--284, [\href{http://arxiv.org/abs/1005.0655}{{\tt arXiv:1005.0655}}].

\bibitem{Sonnenschein:2016pim}
J.~Sonnenschein, {\it {Holography Inspired Stringy Hadrons}},
  \href{http://arxiv.org/abs/1602.00704}{{\tt arXiv:1602.00704}}.

\bibitem{Klebanov:2000hb}
I.~R. Klebanov and M.~J. Strassler, {\it {Supergravity and a confining gauge
  theory: Duality cascades and chi SB resolution of naked singularities}},
  {\em JHEP} {\bf 08} (2000) 052,
  [\href{http://arxiv.org/abs/hep-th/0007191}{{\tt hep-th/0007191}}].

\bibitem{Krishnan:2008gx}
C.~Krishnan and S.~Kuperstein, {\it {The Mesonic Branch of the Deformed
  Conifold}},  {\em JHEP} {\bf 05} (2008) 072,
  [\href{http://arxiv.org/abs/0802.3674}{{\tt arXiv:0802.3674}}].

\bibitem{Basu:2012ae}
P.~Basu, D.~Das, A.~Ghosh, and L.~A. Pando~Zayas, {\it {Chaos around
  Holographic Regge Trajectories}},  {\em JHEP} {\bf 05} (2012) 077,
  [\href{http://arxiv.org/abs/1201.5634}{{\tt arXiv:1201.5634}}].

\bibitem{Bardakci:1969cs}
K.~Bardakci and H.~Ruegg, {\it {Reggeized resonance model for arbitrary
  production processes}},  {\em Phys. Rev.} {\bf 181} (1969) 1884--1889.

\bibitem{Chan:1969ex}
H.-M. Chan and S.~T. Tsou, {\it {Explicit construction of the n-point function
  in the generalized veneziano model}},  {\em Phys. Lett.} {\bf B28} (1969)
  485--488.

\bibitem{DiVecchia:2007vd}
P.~Di~Vecchia, {\it {The Birth of string theory}},  {\em Lect. Notes Phys.}
  {\bf 737} (2008) 59--118, [\href{http://arxiv.org/abs/0704.0101}{{\tt
  arXiv:0704.0101}}].

\bibitem{Koba:1969kh}
Z.~Koba and H.~B. Nielsen, {\it {Manifestly crossing invariant parametrization
  of n meson amplitude}},  {\em Nucl. Phys.} {\bf B12} (1969) 517--536.

\end{thebibliography}\endgroup
\end{document}